\documentclass[prb,twocolumn,showpacs,amsmath,amssymb,longbibliography]{revtex4}

\usepackage{graphicx}
\usepackage{dcolumn}
\usepackage{bm}
\usepackage{gensymb}
\usepackage{multirow}
\usepackage[none]{hyphenat}
\usepackage{epstopdf}
\usepackage{float}
\usepackage{subfigure}
\usepackage{tabularx}
\usepackage{color}
\usepackage{siunitx}
\usepackage{xr}
\usepackage{amsmath} 
\usepackage{mathtools}
\usepackage[normalem]{ulem}
\usepackage{natbib}
\usepackage[caption=false]{subfig}
\usepackage{afterpage}
\usepackage{enumerate}
\makeatletter
\newcommand*{\rom}[1]{\expandafter\@slowromancap\romannumeral #1@}
\makeatother

\begin{document}

\newcommand{\vl}[1]{{\color{blue} #1}}
\newcommand{\beq}{\begin{equation}}
\newcommand{\eeq}{\end{equation}}
\newcommand{\vS}{\vec{S}}
\newcommand{\ud}{\mathrm{d}}


\title{Topological Weyl magnons and thermal Hall effect in layered honeycomb ferromagnets}

\author{Shuyi Li}
\affiliation{Department of Physics and Astronomy, Rice University, Houston, TX 77005, USA}

\author{Andriy H. Nevidomskyy}
\email[Correspondence e-mail address: ]{nevidomskyy@rice.edu}
\affiliation{Department of Physics and Astronomy, Rice University, Houston, TX 77005, USA}

\date{\today}

\begin{abstract}
In this work, we study the topological properties and magnon Hall effect of a  three-dimensional ferromagnet in the ABC stacking honeycomb lattice, motivated by the recent inelastic neutron scattering study of CrI$_3$. We show that the magnon band structure and Chern numbers of the magnon branches are significantly affected by the interlayer coupling $J_c$, which moreover has a qualitatively different effect in the ABC stacking compared to the AA stacking adopted by other authors. The nontrivial Chern number of the lowest magnon band is stabilized by the next-nearest-neighbour Dzyaloshinskii-Moriya interaction  in each honeycomb layer, resulting in the hopping term similar to that in the electronic Haldane model for graphene. However, we also find several gapless Weyl points, separating the non-equivalent Chern insulating phases, tuned by the ratio of the interlayer coupling $J_c$ and the third-neighbour Heisenberg interaction $J_3$. We further show that the topological character of magnon bands results in non-zero thermal Hall conductivity, whose sign and magnitude depend on $J_c$ and the intra-layer couplings. Since the interlayer coupling strength $J_c$ can be easily tuned by applying pressure to the quasi-2D material such as CrI$_3$, this provides a potential route to tuning the magnon thermal Hall effect in an experiment.     
\end{abstract}


\maketitle

\section{Introduction}
Magnons, the low-energy collective excitations of interacting localized spins, serve as the elemental magnetic carrier in insulating magnets~\cite{magnon1,magnon2,magnon3}. Magnons have been demonstrated to form a macroscopic coherent state by quasiequilibrium Bose--Einstein condensation, and can propagate spin information much further than spin current in metals~\cite{magnon4,magnon5,magnon6,magnon7,magnon8,magnon9}. For its potential applications in spintronics field, the topological nature and transport properties of magnons in quantum materials has been one of the subjects of intense interest.
Motivated by the Dirac dispersion of electron states in graphene, linear crossings of magnon bands in honeycomb ferromagnets have been called ``Dirac magnons"~\cite{diracmg1,diracmg2}. 
In analogy with the electronic Haldane model~\cite{Haldane}, gapping out these Dirac points can result in nonzero Chern number of the magnon bands~\cite{owerre1,owerre2,honey1}, achieved by introducing the second neighbor Dzyaloshinskii-Moriya (DM) interaction on the honeycomb lattice ~\cite{DM1,DM2}. The topological nature of the magnon bands in turn results in a nontrivial contribution to the thermal Hall effect~~\cite{owerre1,owerre2,honey1}.
In analogy with the quantum Hall effect of electrons, the magnon thermal Hall effect has thus become one of the most fascinating phenomena with a series of recent theoretical~\cite{MagnonHEth1,Onose297,MagnonHEth2,MagnonHEth3,MagnonHEth4,MagnonHEth5,MagnonHEth6} and experimental~\cite{Onose297,kagome1,kagome2,kagome3,kagome4,lieb1} studies. 

The magnon thermal Hall effect was first predicted theoretically~\cite{MagnonHEth1} in the kagome and pyrochlore ferromagnets with a nearest-neighbor (NN) Dzyaloshinskii-Moriya (DM) interaction~\cite{DM1,DM2} and discovered experimentally~\cite{Onose297} in Lu$_2$V$_2$O$_7$. Subsequently, it has been found that the motion of magnon wave packet along the edge is responsible for the magnon Hall effect with the analytical relation between its magnitude and the Berry curvature of magnon bands~\cite{MagnonHEth2,MagnonHEth3,Murakami2016}. Two-dimensional (2D) magnon thermal Hall effect has been studied in detail on several lattices, including kagome~\cite{MagnonHEth1,kagome1,kagome2,kagome3,kagome4}, Lieb~\cite{lieb1} and honeycomb~\cite{owerre1,owerre2,honey1}. 
However, there has been comparatively little study to date of the magnon thermal Hall effect in the three-dimensional (3D) case.

As we shall demonstrate in this work, one can view the 3D topological magnons in a layered honeycomb ferromagnet as a bosonic analog of 3D electronic topological Weyl semimetals, which have been an active research area~(see e.g. Refs.~[\onlinecite{Weyl-Wan2011, Weyl-Chern-Xu2011, Weyl-Chern-Yang2011,EFMWeylSM1}] for early works, as well as a review~[\onlinecite{Weyl-felser}] and references therein). In electronic Weyl semimetal with broken time-reversal symmetry, the pair of Weyl points can in principle be moved in the Brillouin zone, resulting in either an ordinary insulator (when the Weyl points merge) or in a 3D quantum anomalous Hall insulator (when the Chern band is fully occupied)~\cite{EFMWeylSM1,EFMWeylSM2}. Thus, one can view Weyl semimetal as an intermediate gapless phase between these two insulating phases. In this paper, we show that an analogous intermediate gapless phase, this time not electronic but magnonic in nature, appears naturally in honeycomb ferromagnets with spin-orbit coupling, separating two topological insulating phases with different Chern numbers of the magnon bands. We refer to this gapless phase as a \textit{topological Weyl magnon conductor}.  We note that while the appearance of Weyl magnon points has been addressed in previous theoretical works motivated largely by pyrochlore frustrated magnets~\cite{weyl1,weyl2,weyl3,weyl4,weyl5,weyl6,weyl7,weyl8,weyl9,weyl11,weyl10,weyl12,weyl13}, the emphasis was rather on the bulk-boundary correspondence and the manifestations of the chiral anomaly under the application of the electric field gradient~\cite{weyl9, weyl11}. 
In this work, by contrast, we focus on the effect of the Weyl magnons on the thermal Hall conductivity and in particular formulate the appearance of the gapless Weyl phase as an intermediate phase between two magnon Chern insulators.    
This finding motivates the search for different topological phases and phase transitions between them in various 3D magnetic insulators.
A recent experimental observation of topological Dirac magnons in a 3D collinear antiferromagnet Cu$_3$TeO$_6$~\cite{weyl5,weyl6} may serve as an experimental platform for observing the topological Weyl magnons.

A recent inelastic neutron scattering on CrI$_3$ characterized the spin-wave excitations in this material, with the indication that low-lying magnon bands may be topological~\cite{cri3}. 
The bulk  CrI$_3$ undergoes ferromagnetic ordering of localized Cr spins below the Curie temperature $T_c=61$~K, with spins oriented along the easy $c$-axis \cite{cri3structure1,cri3structure2,cri3structure3}.
The lattice structure is ABC-stacked honeycomb lattice (see Fig.~\ref{abcs}) and the experimental determined parameters show that there is a non-negligible interlayer coupling $J_c$, as well as the 3rd neighbor intra-layer interaction $J_3$. We note that previous theoretical work on 3D honeycomb lattice~\cite{weyl11,owerre3d} has adopted  the AA stacking instead. As we shall demonstrate, the ABC stacking leads to qualitatively different conclusions regarding the topological properties of the magnon bands and the associated thermal Hall effect, directly applicable to CrI$_3$.

In this work, we study the topological properties and magnon thermal Hall effect of 3D insulating ferromagnets with ABC-stacked honeycomb planes. Focusing on the case of CrI$_3$ for concreteness, we adopt a three-dimensional spin-$3/2$ Heisenberg model with 2nd nearest-neighbor DM interactions, using the experimentally determined exchange constants~\cite{cri3}. We show that the interlayer coupling $J_c$ and the 3rd NN Heisenberg interaction $J_3$ lead to several different gapped and gapless phase, depending on the ratio of $J_3/J_c$. Notably, we find \textit{topological Weyl magnon conductor} sandwiched between two Chern magnon insulating phases. 
We obtain the analytical formula of thermal Hall conductivity $\kappa_{xy}$ in two limits of very low and very high temperature, both of which are accessible in CrI$_3$ due to the relatively low value of the exchange couplings (of the order of $J_1 \sim 2$~meV). Importantly, we demonstrate that the phenomenology of thermal Hall response in a magnon Weyl conductor is qualitatively different from the fermionic Weyl semimetal in that $\kappa_{xy}$ does not scale linearly with the distance between the magnon Weyl points at any realistic temperature (except at very high temperatures of the order of magnon bandwidth). This is to be juxtaposed with electronic Weyl semimetals where the Hall response at low temperatures is linear in the separation between the Weyl points~\cite{Weyl-Chern-Xu2011, Weyl-Chern-Yang2011}.

We further investigate the sign change of the magnon thermal conductivity, which we show could be used to infer information about the topology of magnon bands.
In the low temperature limit, we observe the sign switch of  the thermal Hall effect upon varying the 3rd neighbor coupling, explained by the sign switch of Berry curvature at the $\Gamma$ point. This analytical finding is corroborated by numerical calculations of $\kappa_{xy}$ over a wide temperature range.
Intriguingly, in a certain parameter regime, we also find  the sign change of  the thermal Hall effect upon varying the interlayer coupling $J_c$, opening up the possibility of a uniaxial stress-induced control of $\kappa_{xy}$. We believe that this provides a route to tuning the magnon thermal Hall effect in CrI$_3$ and related layered materials such as CrGeTe$_3$~\cite{CGT-Lin2017}, and our results offer new guidance for experiments.     

This paper is organized as follows. In Sec.~\ref{2}, we introduce the spin model of ABC-stacked honeycomb lattice and its representation in the linear spin wave theory. The boundary of a \textit{gapless Weyl magnon} phase is established, alongside the Chern insulating phases. In Sec.~\ref{3}, we compute the Berry curvature, Chern number,  and their behavior in the gapped and gapless phases. In Sec.~\ref{4}, we derive the analytical results of thermal Hall $\kappa_{xy}$ at low and hign temperature limit, complemented with numerical calculations across the entire temperature range.  We identify the sign change of thermal conductivity upon varying $J_3$ and $J_c$ in Sec.~\ref{5}, and draw conclusions.   

\section{Lattice structure and spin model}\label{2}

\subsection{CrI$_3$ lattice structure and Hamiltonian}
\begin{figure}[H]
    \centering
    \includegraphics[width=0.3\textwidth]{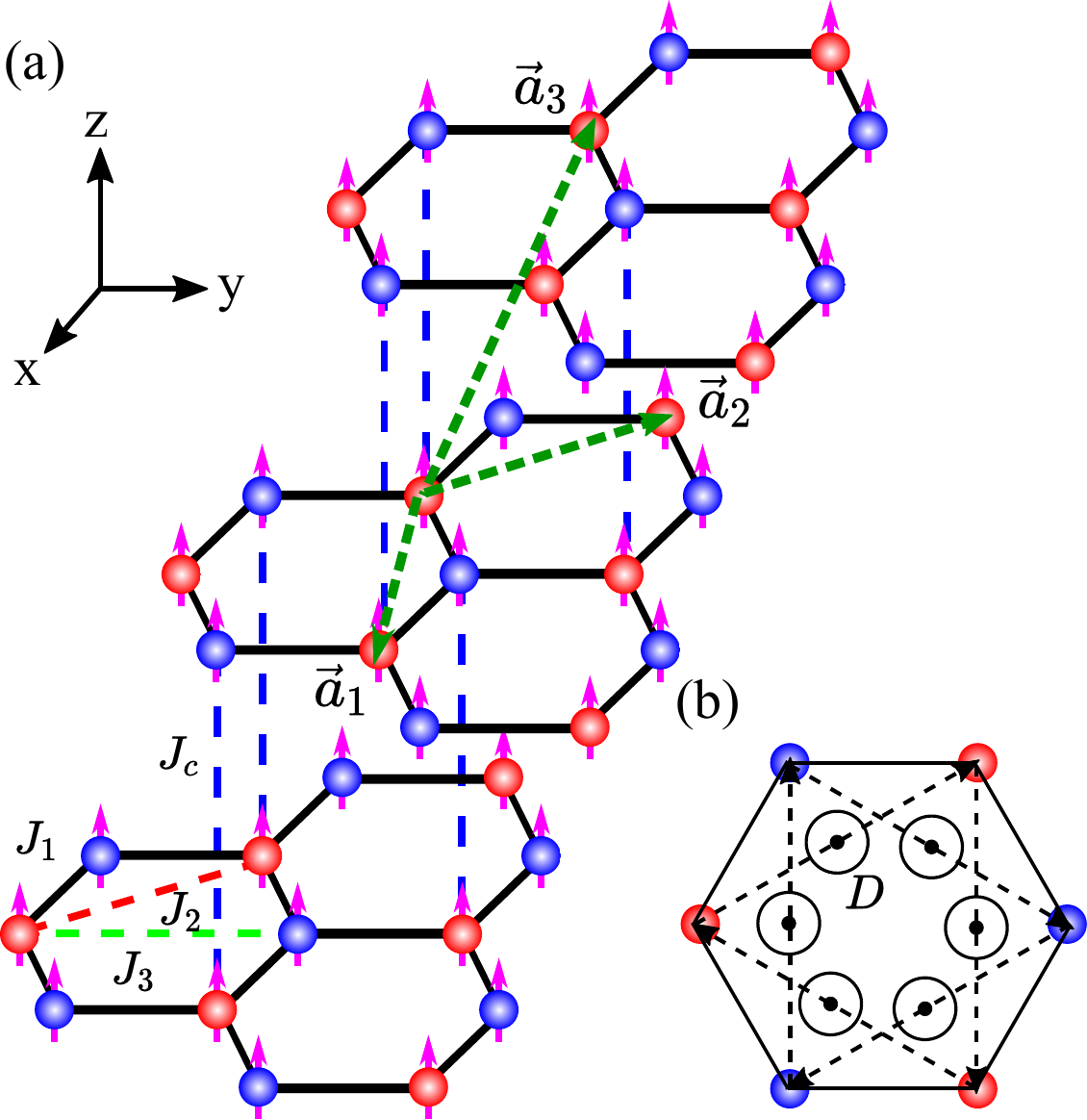}
    \caption{(a) Lattice structure of CrI$_3$, in one layer the 1st, 2nd and 3rd nearest neighbor Heisenberg interactions are $J_1$ (black), $J_2$ and (red) $J_3$ (green), $J_c$ (blue) is the interlayer coupling. $\vec{a}_1$, $\vec{a}_2$ and $\vec{a}_3$ are primitive translation vectors; (b) DM interaction $D$ between next nearest neighbor.}
    \label{abcs}
\end{figure}

The ferromagnetism of CrI$_3$ is due to Cr$^{3+}$ ions, which form a network of honeycomb layers. The layers are stacked against each other by van der Waals interactions, with the structure becoming rhombohedral  (space group $R\bar{3}$, no. 148) below 90~K. The lattice structure can be approximated as ABC stacking of honeycomb layers in the $z$ direction, as  shown in Fig.~\ref{abcs}(a). The primitive translation vectors are $\vec{a}_1=(\sqrt{3}a,0,0)$, $\vec{a}_2=(-\sqrt{3}a/2,3a/2,0)$ and $\vec{a}_3=(0,a,c)$. 
For simplicity, we set $a=c=1$ in what follows as this does not affect qualitatively our conclusions.

We use $S=\frac{3}{2}$ Heisenberg Hamiltonian to model this system, following the inelastic neutron scattering study where the model parameters have been determined by fitting the linear spin-wave spectra~\cite{cri3}:
\begin{equation}\label{eq.hamiltonian}
\begin{aligned}
    H&=-\sum_{\langle ij\rangle,l}J_{ij}\vec{S_i}^l\cdot\vec{S_j}^l+\sum_{\langle\langle ij\rangle\rangle,l}\vec{D}_{ij}\cdot(\vec{S_i}^l\times\vec{S_j}^l)\\
    &-J_c\sum_{\langle ll'\rangle,\langle ij\rangle}\vec{S_i}^l\cdot\vec{S_j}^{l'}-K\sum_{i,l}(S_i^{l,z})^2,
\end{aligned}
\end{equation}
where indices $l$ and $l'$ run over the layers and $J_{ij}$ describes the intra-layer interactions: first, second and third nearest neighbor coupling, $J_1$, $J_2$ and $J_3$.  The second term is the Dzyaloshinskii-Moriya interaction, with the bond-dependent $\vec{D}_{ij}$ vectors determined by Moriya's rules: $\vec{D}_{ij}=Dv_{ij}\,\hat{z}$, with $v_{ij}=+1$ ($-1$) for clockwise (anti-clockwise) direction, respectively, as shown in Fig.~\ref{abcs}(b). The third term describes the interlayer nearest neighbor coupling $J_c$ between adjacent layers, and the last term captures the single-ion Ising anisotropy responsible for the easy axis of Cr$^{3+}$ spins. Previous theoretical and experimental studies have established that the DM interaction between next-nearest neighbors leads to the topological Chern magnon bands in a 2D honeycomb lattice, resulting in a  non-trivial thermal hall effect~\cite{owerre1,owerre2}. 

\subsection{Linear spin wave expansion}
We analyze the spin Hamiltonian in Eq.~(\ref{eq.hamiltonian}) using the linear spin wave expansion. Owing to the single-ion Ising anisotropy on Cr$^{3+}$ site, we choose the $z$ direction to be along the easy axis (parallel to the crystallographic $c$ axis). Then, the spin operators are expressed using the standard Holstein--Primakoff transformation:
\begin{equation}\label{eq.HP}
\begin{aligned}
S_i^+&=\sqrt{2S-a_i^{\dagger}a_i}a_i,\\
S_i^-&=a_i^{\dagger}\sqrt{2S-a_i^{\dagger}a_i}\\
S_i^z&=S-a_i^{\dagger}a_i.
\end{aligned}
\end{equation}
with $S=\frac{3}{2}$. 
 After transforming into momentum space and retaining only bilinears of bosons, the Hamiltonian can be expressed as $H=\sum_{\vec{k}}\mathbf{b}_{\vec{k}}^{\dagger}\cdot \mathcal{H}(\vec{k})\cdot \mathbf{b}_{\vec{k}}$, where $\mathbf{b}_{\vec{k}}^{\dagger}=(b_{A\vec{k}}^{\dagger},b_{B\vec{k}}^{\dagger})$. The Hamiltonian matrix can be written succinctly as 
\begin{equation}\label{eq.matrixH}
\mathcal{H}(\vec{k})=h_0(\vec{k})\sigma_0+h_x(\vec{k})\sigma_x+h_y(\vec{k})\sigma_y+h_z(\vec{k})\sigma_z,
\end{equation}
where  $\sigma_i$'s are the Pauli matrices in the sublattice (A,B) space. The explicit expressions for the coefficients $h_i(\vec{k})$ can be found in Appendix~\ref{app.H}.

\begin{figure}[H]
    \includegraphics[width=0.45\textwidth]{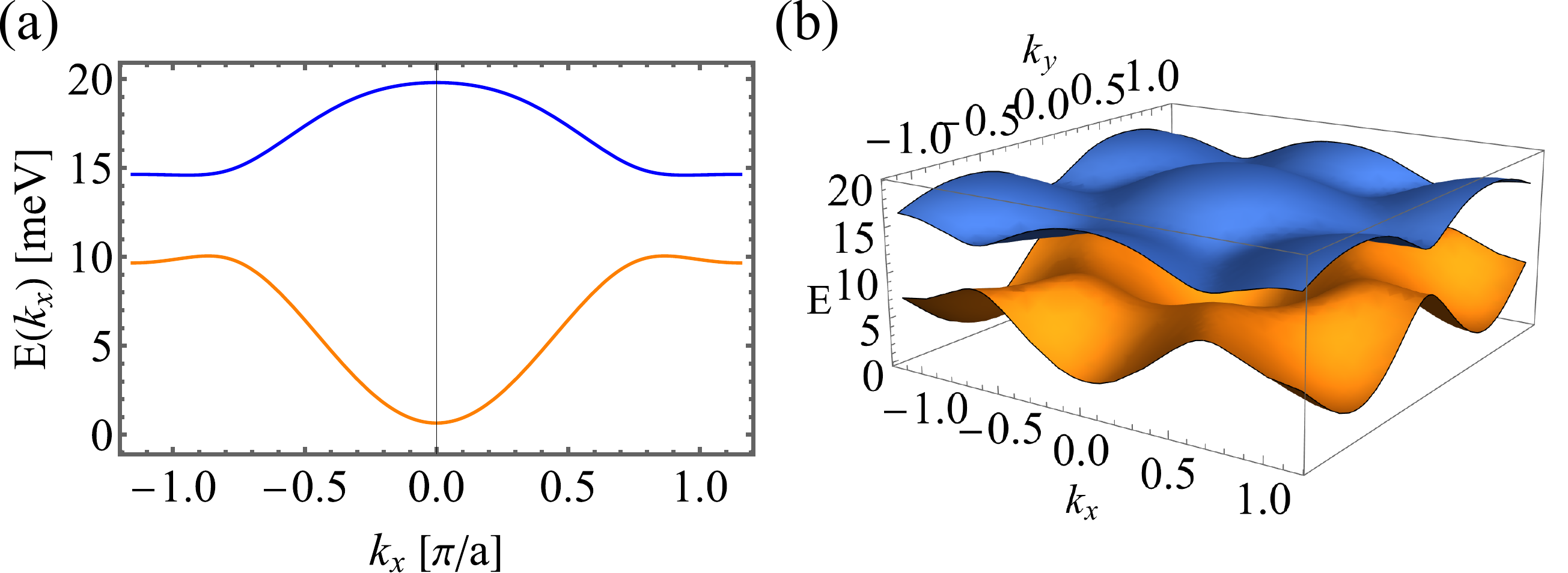}
    \caption{Band structure with experiment fitting parameters $J_1=2.01$ meV, $J_2=0.16$ meV, $J_3=-0.08$meV, $J_c=0.59$ meV, $D=0.31$ meV, $K=0.22$ meV. (a) 2D plot of two bands for $k_y=0$, $k_z=0$, (b) 3D plot of two bands at $k_z=0$.}
    \label{23dband}
\end{figure}

The resulting energy spectrum is given by
\begin{equation}\label{energy}
\epsilon_{\lambda}(\vec{k})=h_0(\vec{k})+\lambda\sqrt{h_x(\vec{k})^2+h_y(\vec{k})^2+h_z(\vec{k})^2},
\end{equation}
where $\lambda$ is $-1$ ($+1$) for the lower (upper) magnon band, respectively. The resulting band structure obtained with the experimentally detetemined fitting parameters~\cite{cri3} is shown in Fig.\ref{23dband}.
The band gap in the magnon spectrum is given by
\begin{equation}
\Delta\epsilon(\vec{k})=2\sqrt{h_x(\vec{k})^2+h_y(\vec{k})^2+h_z(\vec{k})^2}.
\end{equation}

\subsection{Parameter range of gapped and gapless phases}

The Dzyaloshinskii--Moriya interaction, which breaks inversion symmetry, is the key factor in endowing the magnon bands in honeycomb ferromagnets with nontrivial topology~\cite{owerre1}. In the absence of the DM interaction, the magnon energy spectrum is gapless. The presence of arbitrarily small 2nd-neighbor DM interaction $D\neq0$ results in a complex phase factor to the corresponding  magnon hopping term, in direct analogy to the electronic Haldane model~\cite{Haldane}, and opens up a gap in the spectrum of 2D honeycomb model relevant for a monolayer CrI$_3$. It is well established that the resulting magnon bands have a nontrivial Chern number~\cite{owerre1,owerre2}. 
However, as we demonstrate below, the presence of the interlayer spin coupling $J_c$ between the ABC-stacked layers results in a number of topologically non-trivial phases, some of them gapless and some of them retaining the gap in the magnon spectrum. 

Consider first the condition for the gapless spectrum in Eq.~(\ref{energy}), which translates into three equations $h_x(\vec{k})=h_y(\vec{k})=h_z(\vec{k})=0$. For the purely 2D monolayer model, the gapless condition is only sarisfied at an isolated point $J_3/J_1=1/3$. However when interlayer coupling $J_c$ is introduced, the gapless phase is found for a range of $J_3$ values (see Appendix~\ref{app.H}):
\begin{equation}\label{J3range1}
\frac{1}{3}-\frac{|J_c|}{3J_1}<\frac{J_3}{J_1}<\frac{1}{3}+\frac{|J_c|}{3J_1},
\end{equation}
provided the interlayer coupling is not too large: $0<\frac{J_c}{J_1}\le\frac{1}{8}(-5+3\sqrt{17})\approx0.921$. 
This gapless phase is flanked on either side by fully gapped magnon insulators which, as we show in the next section, have nontrivial Chern numbers.
We note paranthetically that the model also admits other gapless solutions, however they correspond to unphysically large values of $J_c$ or $J_3$ and are therefore discarded in what follows. 

\section{Berry curvature and Chern number}\label{3}
The momentum space Berry curvature is given by the standard expression
\begin{equation}\label{eq.BC}
    \Omega_{\alpha\beta}^n(\vec{k})=-2\text{Im}\sum_{m\neq 
    n}\frac{\left\langle P_{\vec{k}n}|\hat{v}_{\alpha}|P_{\vec{k}m}\right \rangle\left\langle P_{\vec{k}m}|\hat{v}_{\beta}|P_{\vec{k}n}\right\rangle}{(\epsilon_n(\vec{k})-\epsilon_m(\vec{k}))^2},
\end{equation}
where $\hat{v}_{\alpha}=\partial\mathcal{H}(\vec{k})/\partial k_{\alpha}$ and the pair of indices $\alpha \neq\beta=\{x,y,z\}$ define a set of two-dimensional planes in $k$-space in which the Berry curvature is computed. $P_{\vec{k}n}$ is the eigenvector corresponding to the eigenvalue $\epsilon_n(\vec{k})$ of matrix $\mathcal{H}(\vec{k})$. In the ferromagnetic case, Eq.(\ref{eq.BC}) can be transformed (see e.g. Ref.~[\onlinecite{owerre1,owerre2}])
\begin{equation}\label{eq.BC1}
    \Omega_{\alpha\beta}^n(\vec{k})=-2\text{Im}\left\langle\frac{\partial}{\partial k_{\alpha}}P_{\vec{k}n}|\frac{\partial}{\partial k_{\beta}}P_{\vec{k}n}\right\rangle.
\end{equation}

For the layered system such as CrI$_3$, one can consider them as a stack of two-dimensional $xy$-planes, for which the Chern number can be computed for any fixed value of $k_z \in [0, 2\pi]$ as an integral of the corresponding Berry curvature over the 2D section of the Brillouin zone:
\begin{figure}[htbp]
    \includegraphics[width=0.25\textwidth]{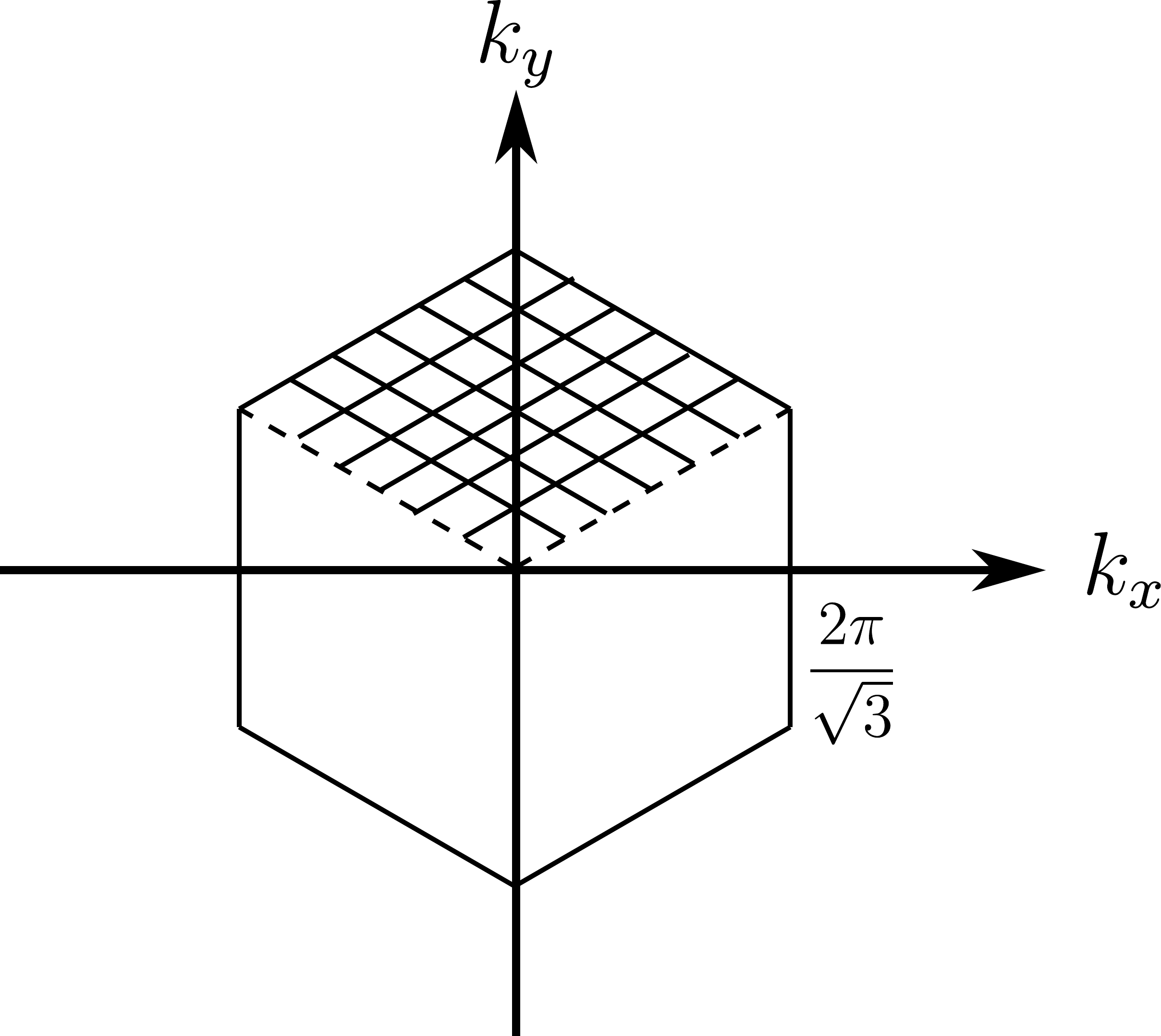}
    \caption{{The minimum translationally invariant region in reciprocal space parallel to $kx$-$ky$ plane, which has an area of three times that of the first 2D Brillouin zone. By the rotational symmetry, the integral over the dashed area is equivalent to the integral over the first Brillouin zone in Eq.~(\ref{eq.Chern}).
    }}
    \label{effBZ}
\end{figure}
\begin{equation}
    \mathcal{C}_{\lambda}(k_z)=\frac{1}{2\pi}\int_{BZ}d\vec{k}_{\parallel}\ \Omega_{xy}^{\lambda}(\vec{k}_{\parallel},k_z),
    \label{eq.Chern}
\end{equation}
where the index $\lambda=\pm 1$ denotes the two magnon bands given by Eq.~(\ref{energy}). \textcolor{black}{In the case of ABC stacking, the minimum translation invariant region in reciprocal space parallel to $(k_x ,k_y)$ plane is a hexagon shown in Fig.~\ref{effBZ}. It contains the area three times that of the first 2D brillouin zone. After considering $120^{\circ}$ rotation symmetry, the 2D integration over $d k_\parallel$ (at a fixed value of $k_z$) is over the 1st Brillouin zone, equivalent to the dashed area in Fig.~\ref{effBZ}. }

\begin{figure}[htbp]
    \includegraphics[width=0.45\textwidth]{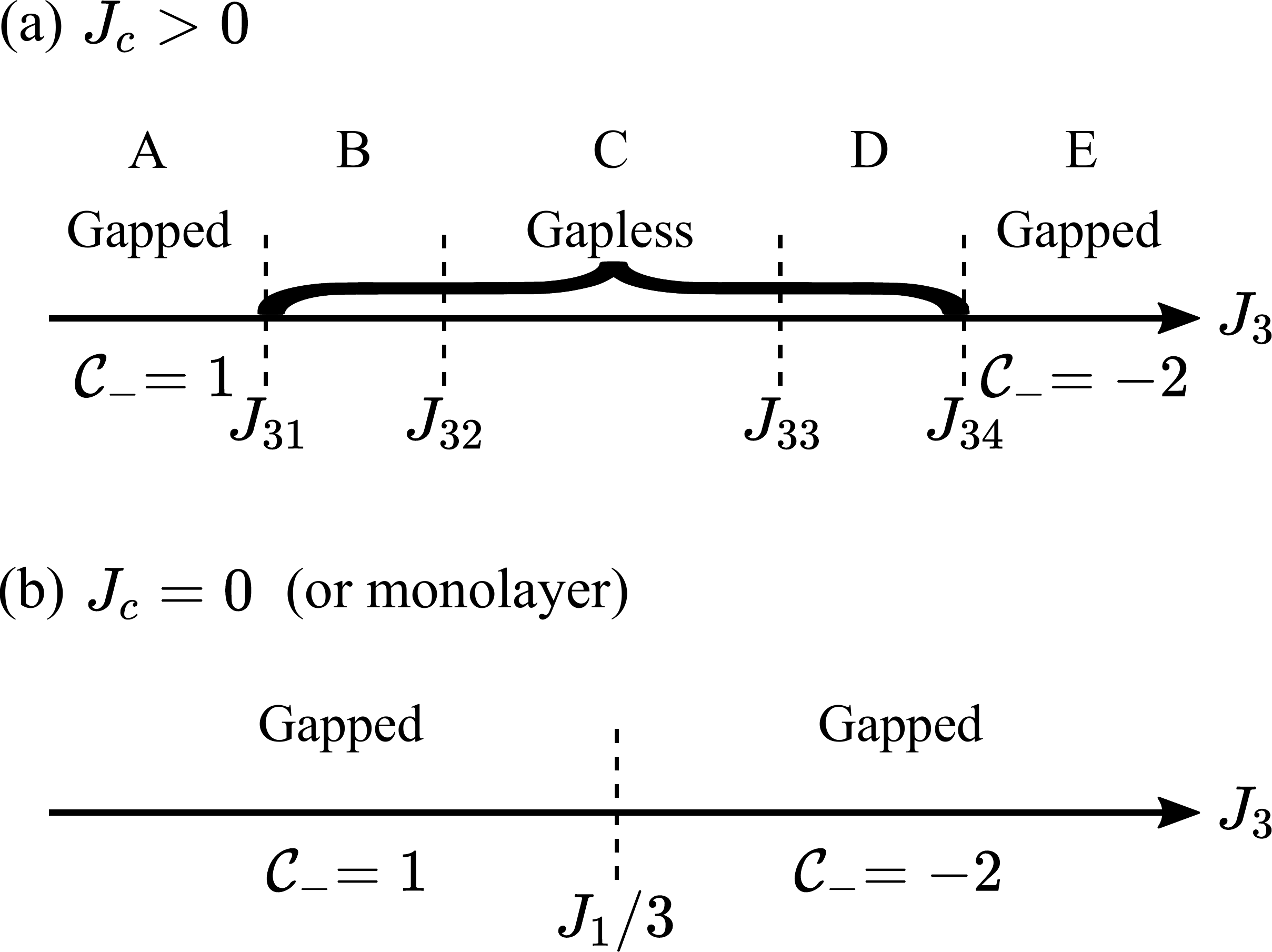}
    \caption{Topological phases and Chern number of the lower magnon band $\mathcal{C}_{-}$ as functions of $J_3$. (a) In the case of finite interlayer coupling $J_c>0$, the gapped phase A has Chern number  $\mathcal{C}_{-}(k_z)=1$ for all $k_z$, whereas phase E has $\mathcal{C}_{-} = -2$.  Points $J_3=J_{31},J_{34}$ mark the boundaries of the gapless phase, where the Weyl points annihilate one another in pairs. The magnon band is gapless in phases B, C, and D, which are Weyl magnon phases with different dependence of planar Chern number $\mathcal{C}_{-}(k_z)$ on $k_z$. At $J_3=J_{32},J_{33}$, two Weyl points with opposite charges cross through the same $k_z$ plane without annihilating each other (see section \ref{sec:gapless} for details). (b) In the case of  vanishing interlayer coupling $J_c=0$ relevant for monolayer, the two gapped phases on either side of $J_{3c}=J_1/3$ have the  same Chern number $C_{-}$ as the phases A and E above, respectively.  }
    \label{cn}
\end{figure}

The integrand can be rewritten in the following convenient form:
\begin{equation}\label{eq.BC2}
  \Omega_{xy}^{\lambda}(\vec{k})=\lambda(\frac{\partial \phi}{\partial k_{y}}\frac{\partial \frac{h_z}{\Delta\epsilon}}{\partial k_{x}}-\frac{\partial \phi}{\partial k_{x}}\frac{\partial \frac{h_z}{\Delta\epsilon}}{\partial k_{y}}),  
\end{equation}
with $\tan \phi=\frac{h_y}{h_x}$. The resulting Chern number for the lower band $\mathcal{C}_{-}$ turns out to depend sensitively on the value of $J_3$ and is shown in Fig.~\ref{cn}(a). The two gapped phases (A and E in Fig.~\ref{cn}(a)) both have non-trivial Chern numbers and are separated by the gapless phase, whose boundaries were derived above in Eq.~(\ref{J3range1}). We note that the gapless Weyl magnon phase exists in a finite range of parameter $J_3$, and three separate phases (B,C and D) can be distinguished, whose nature will be discussed in the section \ref{sec:gapless} below.
By contrast, in the case of the monolayer ($J_c=0$), the gapless phase is limited to a single value of $J_3$, as shown in Fig.~\ref{cn}(b). We analyze all the topological phases in detail below.


\subsection{Gapped phases}
\begin{figure}
    \includegraphics[width=0.48\textwidth]{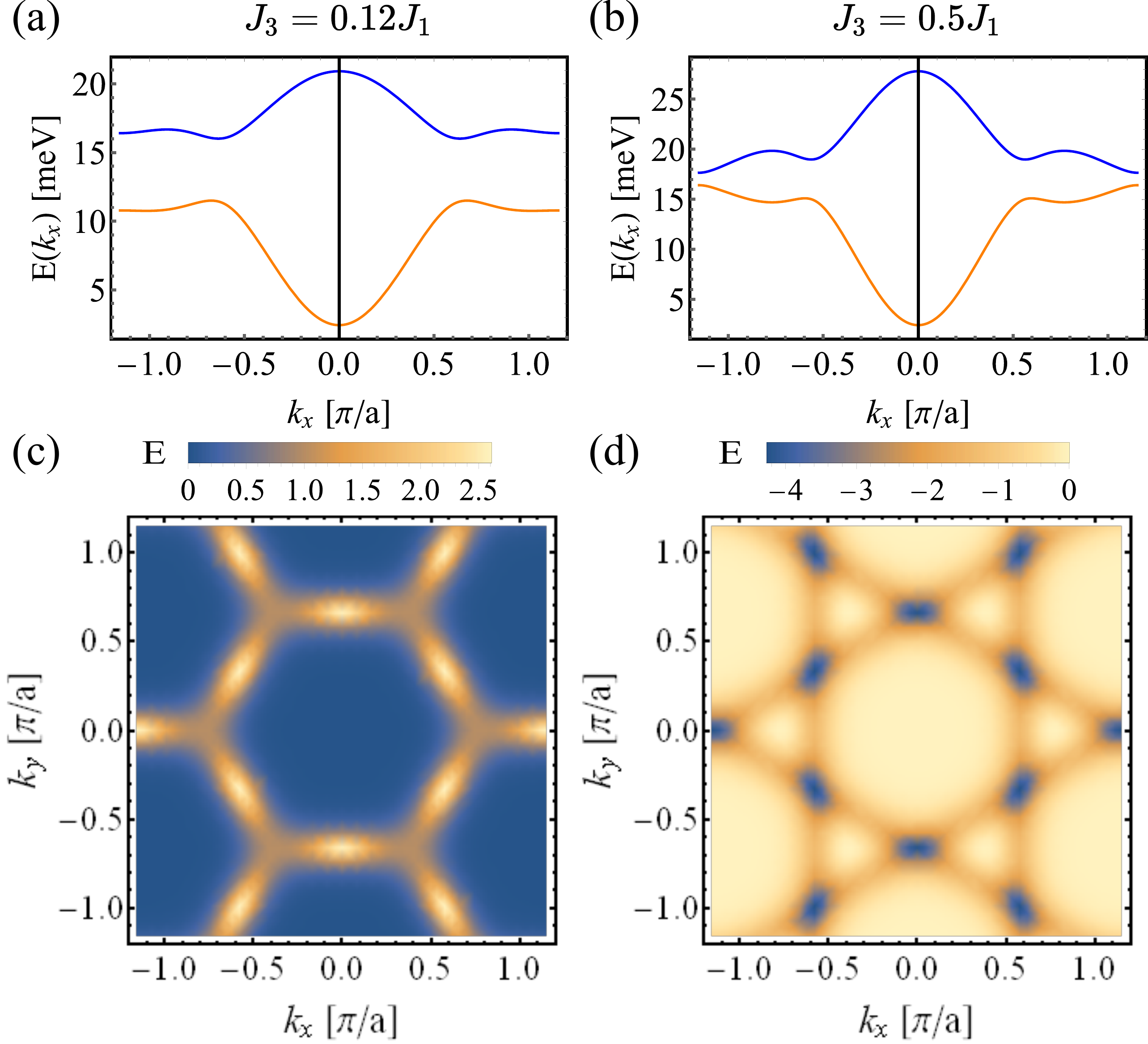}
    \caption{Band structure with experiment fitting parameters $J_1=2.01$ meV, $J_2=0.16$ meV, $J_c=0.59$ meV, $D=0.31$ meV and $K=0.22$ meV. (a) 2D plot of two bands at $k_y=0$, $k_z=\pi$ for $J_3=0.12J_1$. (b) Same as (a) for $J_3=0.5J_1$. (c) (d) Density plot of the Berry curvature for these two cases.}
    \label{gappedBandandBC}
\end{figure}

 In the gapped phases A and E in Fig.~\ref{cn}, the spectral gap between the two magnon bands does not close with varying $k_z$, and the Chern number is well defined and remains the same for all $k_z$. In phase A $J_3<J_{31}=(J_1-J_c)/3$, the Chern number of lower band $\mathcal{C}_{-}=1$. For large values of  $J_3>J_{34}=(J_1+J_c)/3$ in phase E, $\mathcal{C}_{-}=-2$. The magnon band structure and the Berry curvature of these two gapped phases are shown in Fig.\ref{gappedBandandBC}.

\subsection{Gapless phases\label{sec:gapless}}
\begin{figure}[htbp]
    \centering
    \includegraphics[width=0.48\textwidth]{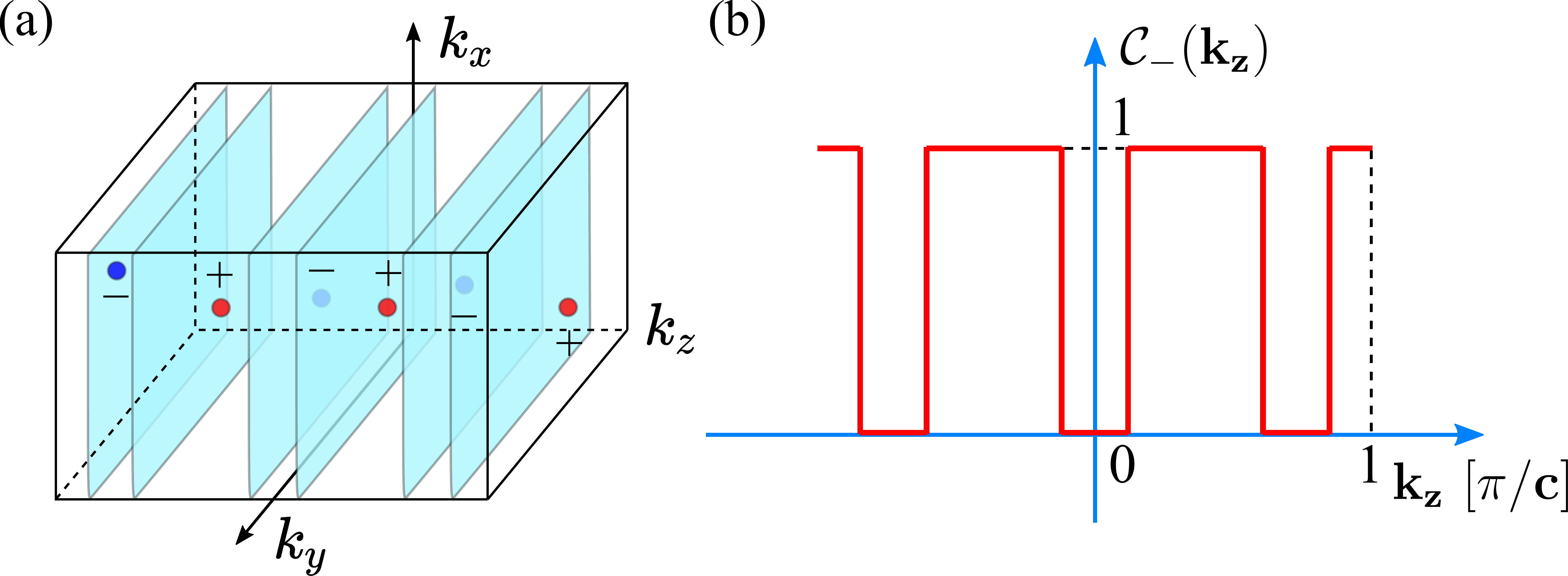}
    \caption{Inside the phase B $J_{31}<J_3<J_{32}$, we sketch (a) Weyl points in $k$ space, (b) the Chern number of lower magnon band $\mathcal{C}_{-}(k_z)$, which varies with $k_z$ and is non-zero only for the planes in-between the pairs of Weyl points.}
    \label{gaplessweylpoint}
\end{figure}

In the gapless phase realized when \mbox{$J_{31} \leq J_3 \leq J_{34}$}, the $xy$ Chern number in Eq.~(\ref{eq.Chern}) is well defined almost everywhere in the Brillouin zone with the exception of six $k_z$ planes which house the monopoles or anti-monopoles of the Berry curvature. These points serve as the sources or sinks of the Berry curvature, with topological charge $+1$ and $-1$, respectively. The behaviour of them with $J_3$ changing can be found in Appendix~\ref{app.Weyl}. In the phases B, C, and D, these six gapless points are localized at six distinct values of $k_z$, and as a result, the Chern number of the lower magnon band $\mathcal{C}_{\pm}(k_z)$ jumps by $\pm 1$ when $k_z$ cross each of these planes. For example, the case of phase B is shown in Fig.~\ref{gaplessweylpoint}. This is in direct analogy with electronic Weyl semimetals, where the Chern number is non-zero in any plane between a pair of Weyl points, and jumps to zero upon crossing the Weyl point  ~\cite{Weyl-Chern-Xu2011, Weyl-Chern-Yang2011}. For this reason, we dub the monopoles of the Berry curvature in the layered ferromagnet \textit{Weyl magnons} and refer to the corresponding gapless phase as a \textit{Weyl magnon conductor}. Of course the original concept of Weyl fermions~\cite{Weyl1929} refers to the solution of the massless Dirac equation in (3+1)D, and the usage of the term  when applied to bosons may be deemed objectionable; nevertheless, the parallels are also very clear -- the role of spin in Weyl spinors is played by the sublattice label (A, B) in the magnon case, the spectrum is linear in both cases, and the Weyl points appear in pairs with equal but opposite chirality, just like in electronic Weyl semimetals.

The discovery of Weyl magnons in layered ferromagnets is one of the key novel results of the present work. This identification raises a natural question -- is there an analog of a transport coefficient in the magnon case, such that one can associate the jump of the Chern number upon crossing the Weyl point with a plateau transition of the corresponding (anomalous) Hall effect, as is the case for electronic Weyl semimetals~\cite{Weyl-Chern-Xu2011, Weyl-Chern-Yang2011}? The answer is ``almost," in a sense that the nontrivial Chern number of the magnon bands results in a generically non-zero value of the thermal Hall conductivity. At the same time, the crucial difference with the electronic case is that the chemical potential for magnons lies at zero energy, meaning that in the limit of zero temperature, the magnon occupation number is zero and the thermal Hall effect vanishes. This difference notwithstanding, there are measurable consequences of the magnon topology at any finite temperature, which we shall analyze now.


\section{Topological thermal hall effect}\label{4}

With the choice of coordinate system where ferromagnetic layers lie perpendicular to the
$z$ direction, as in the convention we have followed, the nontrivial Berry curvature in $(xy)$ plane will result in a nonvanishing value of the transverse (Hall) component of the thermal conductivity $\kappa_{xy}$, which quantifies the generation of a
transverse energy flux $j_x^Q$
upon the application of a temperature gradient $\partial_y T$: $j_x^Q = -\kappa_{xy} \partial_y T$.
The magnon contribution to the $\kappa_{xy}$ is found to be~\cite{MagnonHEth2,MagnonHEth3,MagnonHEth5,Murakami2016}
\begin{equation}\label{kxy}
    \kappa_{xy}=-\frac{k_B^2T}{(2\pi)^3\hbar}\int_{BZ}d\vec{k}\sum_nc_2(f_n(\vec{k}))\Omega_{xy}^n(\vec{k}),
\end{equation}
where $c_2(x)=(1+x)(\ln(\frac{1+x}{x}))^2-(\ln x)^2-2Li_2(-x)$, where $Li_2(x)$ is the polylogarithm function of order 2 and $f_n(\vec{k})=n_B(E_n(\vec{k}))\equiv (e^{E_n(\vec{k})/k_BT}-1)^{-1}$ is the Bose function of magnons at temperature $T$. The minus sign in front of the integral in Eq.~(\ref{kxy}) can be understood from the semiclassical theory in Refs.~[\onlinecite{MagnonHEth2,MagnonHEth3}], which shows that $\kappa_{xy}$ originates from  the magnon edge current, which carries a minus sign relative to the Berry curvature of the band.

One can view the above expression as an integral of the Berry curvature weighted by the  prefactor $c_2(f_n(\vec{k}))$.
It is the energy dependence of this prefactor that makes the value of $\kappa_{xy}$ not quantized, and it is only in the special limit of perfectly flat magnon band $E_n(k)=E_0$ separated by a large gap $\Delta\epsilon \gg k_BT$ that quantization can be achieved~\cite{THE-Loss2017}, up to a constant prefactor $n_B(E_0)$. We note parenthetically that there may be other, non-magnonic contributions to the anomalous thermal Hall effect in insulators, such as for instance due to phonons that are coupled to a chiral quantum spin liquid~\cite{CSL-Savary2018}, however these effects are not subject of the present work and are absent in CrI$_3$ and related layered ferromagnets. 


In general, the expression in Eq.~(\ref{kxy}) must be evaluated numerically, however an analytical solution can be found in two limits: that of very low temperatures compared to the magnon dispersion $k_BT\ll |J|$, and in the opposite limit of very high temperatures. Note that in the case of CrI$_3$ where $J_1/k_B \sim 20$~K, both limits are within experimental reach.
In this section, we first summarize the analytical solutions in these limits, before turning to the numerical analysis at intermediate temperature in the following section. 


\subsection{$k_BT\ll|J|$ limit}
At temperature low compared to the magnon bandwidth $k_BT \ll |J|$, the behavior of the function $c_2(f_n(\vec{k}))\approx (\beta E_n(\vec{k}))^2 e^{-\beta E_n(\vec{k})}$ means that $\kappa_{xy}$ in Eq.~(\ref{kxy}) is dominated by the lowest energy in the magnon dispersion (here $\beta\equiv (k_BT)^{-1}$). The contribution far from the minimum energy point of the lowest magnon band is thus negligibly small. 

We assume that $J_3$ is positive (ferromagnetic in our notation of Eq.~(\ref{eq.hamiltonian})) or if negative, not too large compared to $|J_1|$. In fact, Eq.~(\ref{eq.expansionneargamma}) shows that in the range of $J_3>-\frac{1}{4}J_1-\frac{3}{2}J_2$, $\Gamma=(0,0,0)$ point of the lower band is the minimum energy point, as is the case in CrI$_3$ (see dispersion in Fig.~\ref{23dband}).

Then, expanding the energy and Berry curvature near the $\Gamma$ point, we find
\begin{equation}\label{eq.expansionneargamma}
\begin{aligned}
    \Omega(\vec{k},J_c,\delta)&=-A(J_1,J_3,J_c)\cdot D\cdot\delta\cdot(k_x^2+k_y^2)^2\\
    &+B(J_1,J_3,J_c)\cdot D\cdot (k_y^2-3k_x^2)k_yk_z,\\
    E(\vec{k},J_c,J_3)&=3K+\frac{1}{2m_{\parallel}}(k_x^2+k_y^2)+\frac{1}{2m_z}k_z^2,
\end{aligned}
\end{equation}
where A is a positive function that depends on $J_1$, $J_3$ and $J_c$, $\delta=J_3-J_1/8$, $m_{\parallel}=4/(9(J_1+6J_2+4J_3))$,  $m_z=2(3J_1+3J_3+J_c)/(9(J_1+J_3)J_c)$. The second term in  $\Omega(\vec{k},J_c,\delta)$ is an odd function of $k_z$ and does not contribute when integrating over all $k_z$. Finally, $\kappa_{xy}$ becomes
\begin{equation}\label{T0L}
\begin{aligned}
\kappa_{xy}&=-\frac{k_B^2T}{(2\pi)^3\hbar}\int_{BZ} d^3k\Omega(\vec{k},\delta)\frac{E^2}{(k_BT)^2}\exp\left(-\frac{E}{k_BT}\right)\\
&=\frac{9A K^2}{(2\pi)^3\hbar T}\cdot D\; (J_3 - J_1/8) \,\exp\left(-\frac{3K}{k_BT}\right)\\
&\times (2k_BTm_{\parallel})^3\int d^2{k'}_{\parallel} {k'}_{\parallel}^4\exp(-{k'}_{\parallel}^2)\\ 
&\times \sqrt{2k_BTm_z}\int dk'_z \exp(-{k'}_z^2),
\end{aligned}
\end{equation}
where ${k'}_{\parallel}=k_{\parallel}/\sqrt{2k_Bm_{\parallel}T}$ and ${k'}_z=k_z/\sqrt{2k_Bm_zT}$. In the limit of $T\ll|J|$, $Tm_{\parallel}\rightarrow0$, which allows us to replace the upper limit of the integral over ${k'}_{\parallel}$ by infinity. The result of integration thus becomes a constant independent of $T$. 
As for the integral over $k'_z$, its values can be computed in two limiting cases: $J_c\gg k_BT$ and the monolayer limit $J_c\ll k_BT$. This integral can be written as
\begin{equation}
\begin{aligned}
    \int dk'_z \exp(-{k'}_z^2)&=\frac{\sqrt{\pi}}{2}\ \text{erf}\left(\frac{2\pi}{\sqrt{2k_Bm_zT}}\right)\\
    &=\left\{
\begin{aligned}
&\frac{\sqrt{\pi}}{2},  & J_c \gg k_BT;\\
&\frac{2\pi}{\sqrt{2k_Bm_zT}},  & J_c \ll k_BT,
\end{aligned}
\right.
\end{aligned}
\end{equation}
where erf$(x)$ is the error function.

Collecting all $T$-dependent terms, we conclude that the  temperature dependence in the low temperature limit is
\begin{equation}\label{eq.kxy-low}
\kappa_{xy} \left(k_BT \ll |J|\right) \propto \exp\left(-\frac{3K}{k_BT}\right)\times \left\{
\begin{aligned}
& T^{\frac{5}{2}}, & J_c \gg k_BT;\\
& T^{\frac{3}{2}},  & J_c \ll k_BT,
\end{aligned}
\right.
\end{equation}
In CrI$_3$, $J_c/k_B \sim 6$~K, whereas the Ising anisotropy $K/k_B \sim 2$~K, and one can in principle measure both exponents in Eq.~(\ref{eq.kxy-low}) .

We notice that in Eq.~(\ref{T0L}) the sign of $\kappa_{xy}$ only depends on $J_3$ in $T\ll |J|$ limit. Thus, we expect a sign change of thermal conductivity at a value of $J_3=J_1/8$, which is due to the sign change of the Berry curvature at the $\Gamma$ point.

\subsection{$|J|\ll k_BT < k_BT_c $ limit}
It is interesting to investigate the behavior of thermal conductivity in our model in the high temperature limit compared to the magnon bandwidth, $k_BT \gg |J|$. In the case of CrI$_3$, the magnon bandwidth is about 10~meV (see Fig.~\ref{23dband}), and given the value of the Curie temperature $T_c=61$~K, this limit may not be pertinent, however in other materials with higher $T_c$, it may be worthy of investigation.

\begin{figure}[htbp]
    \hspace*{-5mm}
    \includegraphics[width=0.3\textwidth]{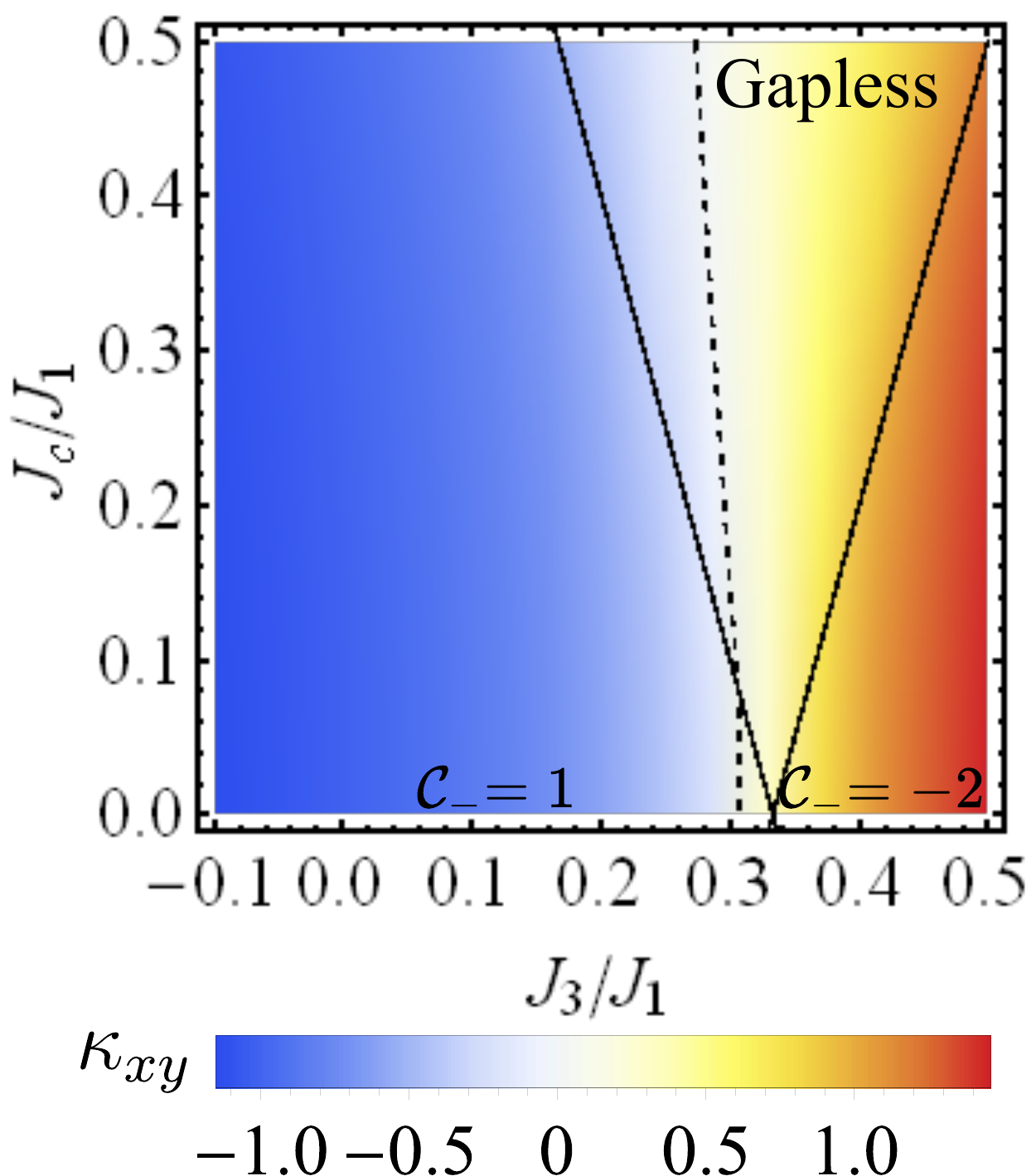}
    \caption{Thermal hall conductivity $\kappa_{xy}$ varying with $J_3$ and $J_c$ in the high temperature limit. The dashed line corresponds to $\kappa_{xy}=0$.}
    \label{HighT}
\end{figure}

At  high temperatures $T\gg|J|/k_B$, the Bose function can be approximated by $f_n(\vec{k})\approx k_BT/E(\vec{k})$. The thermal Hall conductivity  then becomes
\begin{equation}\label{eq.HighT}
\begin{aligned}
    \kappa_{xy}&=-\frac{k_B^2T}{(2\pi)^3\hbar}\int_{BZ}d\vec{k}\sum_n\Omega_n(\vec{k})(-\frac{E_n(\vec{k})}{k_BT})\\
    &=\frac{k_B}{(2\pi)^3\hbar}\int_{BZ}d\vec{k}~\sum_n\Omega_+(\vec{k})\Delta\epsilon(\vec{k}),
\end{aligned}
\end{equation}
which shows that in the high temperature limit, $\kappa_{xy}$ is independent of $T$ and its value depends on the gap $\Delta\epsilon$ between the two branches in the magnon spectrum. Shown in Fig.~\ref{HighT} as a color plot is the behaviour of $\kappa_{xy}$ as a function of couplings $J_3$ and $J_c$.  The dashed line marks the position of zeros $\kappa_{xy}=0$ where it changes sign. 

\subsection{Effect of Weyl points on thermal conductivity}
It was established in Sec.~\ref{3} that a series of topological phase transitions occurs upon increasing the 3rd-neighbour coupling $J_3$, accompanied by the development of Weyl points in the magnon spectrum inside the gapless phase that exists for a range of $J_3$ values: $\frac{J_1 - J_c}{3} < J_3 < \frac{J_1+J_c}{3}$. It is natural to ask how the presence of these Weyl points affects the thermal Hall response. 

\subsubsection{Low temperature regime}
First, in the low-temperature limit, we showed that the value of $\kappa_{xy}(T)$ is entirely determined by the Berry curvature expansion near the lowest-energy point ($\Gamma$) in the lower branch of the magnon spectrum. As such, the closing of the gap between the two branches, which occurs at much higher energies (of the order of $\gtrsim 3 J_1$) and away from the $\Gamma$ points, is not expected to affect the thermal conductivity. Indeed, the analysis of Eq.~(\ref{T0L}) shows that the sign of thermal conductivity is determined by $\mathrm{sgn}(J_3/J_1 - 1/8)$ and does not depend on whether the system is in the gapless regime $\frac{1}{3} - \delta_c <J_3/J_1< \frac{1}{3} + \delta_c$ or not, where we denoted $\delta_c=J_c/(3J_1)$. In other words, the closing of the magnon spectral gap, accompanied by the appearance of the Weyl points, does not affect thermal conductivity in the low-temperature regime $T\ll 4J$  (in the case of CrI$_3$, for $T\lesssim 80$~K, given the magnon bandwidth of $\sim 10$~meV, see Fig.~\ref{23dband}).

Note that this behavior is in stark contrast with that of electronic Weyl semimetals, where the electrical Hall conductance is proportional to the distance in $k$-space between the pairs of Weyl points~\cite{Weyl-Chern-Xu2011, Weyl-Chern-Yang2011}: $\sigma_{xy} \propto \frac{e^2}{h}(|\Delta k|/\pi))$. This is because in the electronic case, the Fermi function replaces the bosonic function $c_2(f(k))$ in the expression for Hall conductivity, and the integration over $k_\parallel$ is performed over all the occupied bands. An alternative way to explain this is that only those values of $k_z$  between the Weyl points are associated with the chiral edge mode in the real-space $xy$ plane that contributes to the electronic Hall conductivity. In the bosonic case, by contrast, even though the chiral edge modes do exist, they are situated at energies of order $J_1$ where the bulk gap opens up in the magnon spectrum, and not at zero energy where the chemical potential for bosons lies. This is why the presence of these chiral modes does not manifest itself in the Hall response until the temperature becomes comparable to the magnon bandwidth, as we shall see shortly.

\subsubsection{Intermediate and high-temperature regime}
At finite temperatures of the order of the magnon bandwidth, one expects both branches of the magnon spectrum to contribute to the thermal conductivity. Indeed, the analysis of the high-temperature regime in Eq.~(\ref{eq.HighT}) manifests that the gap closing $\Delta\epsilon(k_{zi})=0$ at the Weyl points  means that only the set of planes $k_z$ between the pairs of Weyl points contribute to $\kappa_{xy}$. Here, we denoted the $k_z$-momentum positions of Weyl points by $k_z=k_{zi}$, where $i=1...6$.

\begin{figure}[htbp]
    \hspace*{-0.5cm}
    \includegraphics[width=0.5\textwidth]{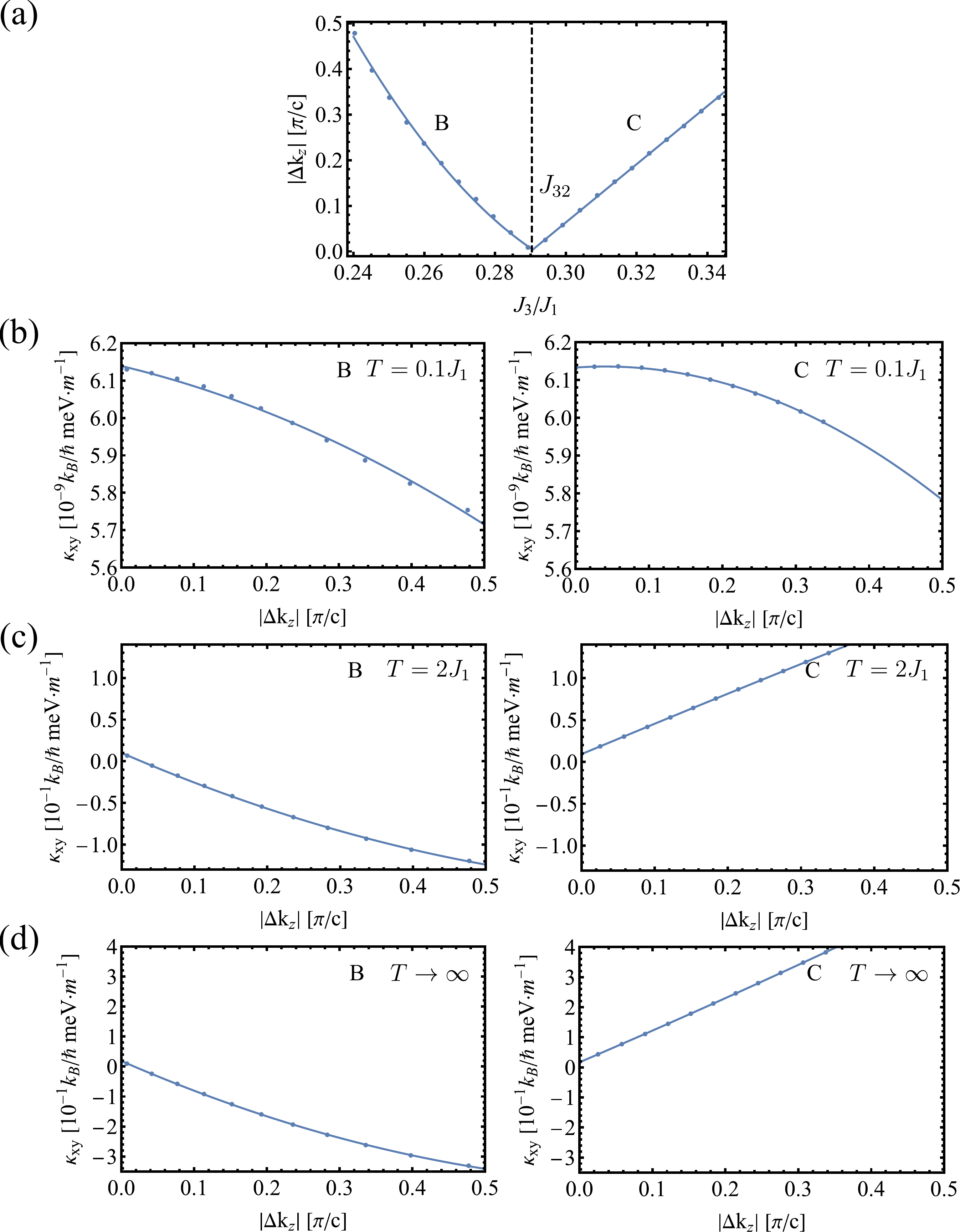}
    \caption{(a) The spacing $|\Delta k_z|$ in $k_z$-space between the planes containing pairs of magnon Weyl points, plotted as a function of the model parameter $J_3/J_1$. In-between the pairs of Weyl points, the Chern number in Eq.~(\ref{eq.Chern}) $\mathcal{C}_{-}(k_z)=+1$ ($-1$) in phase C (phase B), respectively. (b) Thermal hall conductivity $\kappa_{xy}$ plotted against $|\Delta k_z|$ inside the gapless phase B (left panel), and C (right panel) at temperature $T=0.1J1$. (c) Same as (b) at $T=2J_1$. (d) Asymptotic behavior of $\kappa_{xy}$ vs. $|\Delta k_z|$ as $T\to\infty$.}
    \label{kxyvsdk}
\end{figure}

For concreteness, we choose to focus on the gapless phases B and C (see Appendix~\ref{app.Weyl}). The distance between the planes of Weyl points $|\Delta k_z|$ is plotted in Fig.~\ref{kxyvsdk} (a) for these two phases. 
We now plot the thermal conductivity $\kappa_{xy}$ against the Weyl-point separation $|\Delta k_z|$ at three different temperatures $T=0.1J1$, $2J_1$ and $T\to\infty$ in Fig.~\ref{kxyvsdk} (b), (c) and (d). The left (right) panel is for phase B (phase C) respectively.

We fit the data in Fig.~\ref{kxyvsdk} using
\beq
\kappa_{xy}(\Delta k_z, T) = \kappa^{(0)}+\kappa^{(1)}|\Delta  k_z|+\kappa^{(2)}|\Delta k_z|^2
\eeq
for a fixed temperature, and 
the results are summarized in Table~\ref{table1}. 

In phase C, at intermediate temperature $T=2J_1$ and in the high-temperature limit, we find $\kappa^{(1)} \!\gg\! \kappa^{(2)} \gg \kappa^{(0)}$, so that we can  approximate $\kappa_{xy} \propto |\Delta k|$, proportional to the separation between the Weyl points, which is the regime explored in Ref.~\onlinecite{owerre3d} on the example of a stacked kagome antiferromagnet. 
In phase B, by contrast, the behaviour of $\kappa_{xy}(\Delta k_z)$ is more non-linear, although it can still be approximated as roughly \mbox{$\kappa_{xy} \propto |\Delta k|$} at sufficiently high temperatures.

\begin{table}
\caption{Fitting result of $\kappa^{(0)}+\kappa^{(1)}|\Delta k|+\kappa^{(2)}|\Delta k|^2$ in phases B and C at $T=0.1J1$, $T=2J_1$ and $T\to\infty$.  In phase C at $T=2J_1$ and $T\to\infty$, $\kappa^{(1)} \gg \kappa^{(2)} \gg \kappa^{(0)}$, which means that $\kappa_{xy} \propto |\Delta k|$.}
\begin{tabular}{|c||c|c|c|c|}   
\hline 
  Phase & $T$ & $\kappa^{(0)}$ & $\kappa^{(1)}$ & $\kappa^{(2)}$\\
\hline      
  B  &  $0.1J_1$  & $6.14\times10^{-9}$ & $-4.62\times10^{-10}$ & $-7.72\times10^{-10}$\\
\hline 
  C  & $0.1J_1$  & $6.13\times10^{-9}$ & $1.34\times10^{-10}$ & $-1.67\times10^{-9}$\\
\hline      
  B  &  $2J_1$  & $1.00\times10^{-2}$ & $-3.78\times10^{-1}$ & $2.21\times10^{-1}$\\
\hline 
  C  & $2J_1$  & $9.17\times10^{-3}$ & $3.65\times10^{-1}$ & $-2.04\times10^{-2}$\\
\hline      
  B  &  $\gg J_1$  & $1.85\times10^{-2}$ & 1.06 & $6.78\times10^{-1}$\\
\hline 
  C  & $\gg J_1$  & $1.67\times10^{-2}$ & 1.04 & $1.34\times10^{-1}$\\
\hline
\end{tabular}
\label{table1}
\end{table}

\begin{figure}[htbp]
    \hspace*{-0.5cm}
    \includegraphics[width=0.5\textwidth]{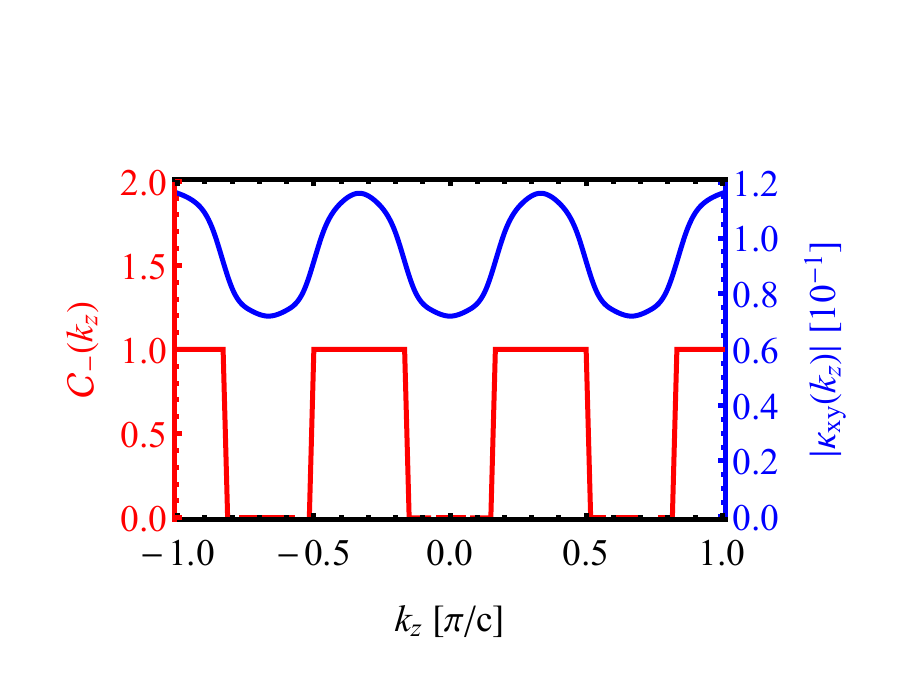}
    \vspace{-10mm}
    \caption{Chern number of the lower band $\mathcal{C}_-(k_z)$ (red) and 2D thermal Hall conductivity $\kappa_{xy}^{2D}(k_z)$ (blue) vs. $k_z$ at $T=2J_1$, $J_c=0.59$ meV and $J_3=J_{31}+J_c/20$ (inside phase B).}
    \label{kxyvskz}
\end{figure}

Returning to the comparison with low-temperature behaviour analyzed in the previous subsection, we also plotted $\kappa_{xy}$ vs. $|\Delta k_z|$ at low temperature $T=0.1J1$  in panel (b) of Fig.~\ref{kxyvsdk}. In that case, thermal conductivity is dominated by the constant term $\kappa^{(0)}$ which originates from the low-energy contribution to Eq.~(\ref{kxy}) near the bottom of the band and does not depends, to first approximation,  on the separation between the Weyl points. In order to see how this picture evolves as temperature is raised, it is useful to write down the thermal conductivity as 
\beq
\kappa_{xy}(T) = \int \frac{\ud k_z}{2\pi} \kappa^{2D}_{xy}(k_z, T),
\eeq
and plot the partial contribution of the 2D quantity $\kappa^{2D}_{xy}$ as a function of $k_z$.  At low temperatures, the dominant contribution comes from near the bottom of the band ($k_z=0$ in CrI$_3$). However at high temperatures, all values of $k_z$ generally contribute (not just between the Weyl planes), as shown in Fig.~\ref{kxyvskz} for $T=2 J_1$. Thus even in the high temperature limit, $\kappa_{xy}$ is not generally expected to scale linearly with $|\Delta k_z|$.


We conclude that the phenomenology of thermal conductivity in a magnon semimetal is drastically different from its fermionic counterpart. 
Nevertheless, the appearance of the Weyl points upon increasing $J_3$ does result in the eventual change of the Chern number of, say, the lower magnon branch from $\mathcal{C}_{-} = +1$ to $\mathcal{C}_{-} = -2$, and it is this change in the sign of the Berry curvature that leads to the sign change of $\kappa_{xy}$ in Eq.~(\ref{eq.HighT}) at high temperatures. Indeed, Fig.~\ref{HighT} shows that the pronounced sign change of thermal conductivity occurs near or in the gapless regime $\frac{1}{3} - \delta_c <J_3/J_1< \frac{1}{3} + \delta_c$ (denoted by solid lines in Fig.~\ref{HighT}) where the Weyl points appear. 

Having established the two limits of low and high temperature, respectively, we now investigate the  behavior and sign change of thermal Hall conductivity at intermediate range of temperatures.


\section{Sign change and temperature dependence of thermal conductivity}\label{5}
As we have already remarked in the previous section, the thermal conductivity changes sign as a function of $J_3$ in both the low- and high-temperature limits. 
In this section, we investigate the sign change behavior further and show the numerical results of evaluating $\kappa_{xy}$ in Eq.~(\ref{kxy}) with varying temperature and coupling constants $J_3$ and $J_c$. For concreteness, we set all the other coupling constants equal to the experimentally determined values~\cite{cri3} for CrI$_3$: $J_1=2.01$ meV, $J_2=0.16$ meV, $K=0.22$ meV and $D=0.31$ meV.      

\subsection{Temperature evolution of $\kappa_{xy}$ as a function of $J_3$}

\begin{figure}[tbp]
    \centering
    \includegraphics[width=0.5\textwidth]{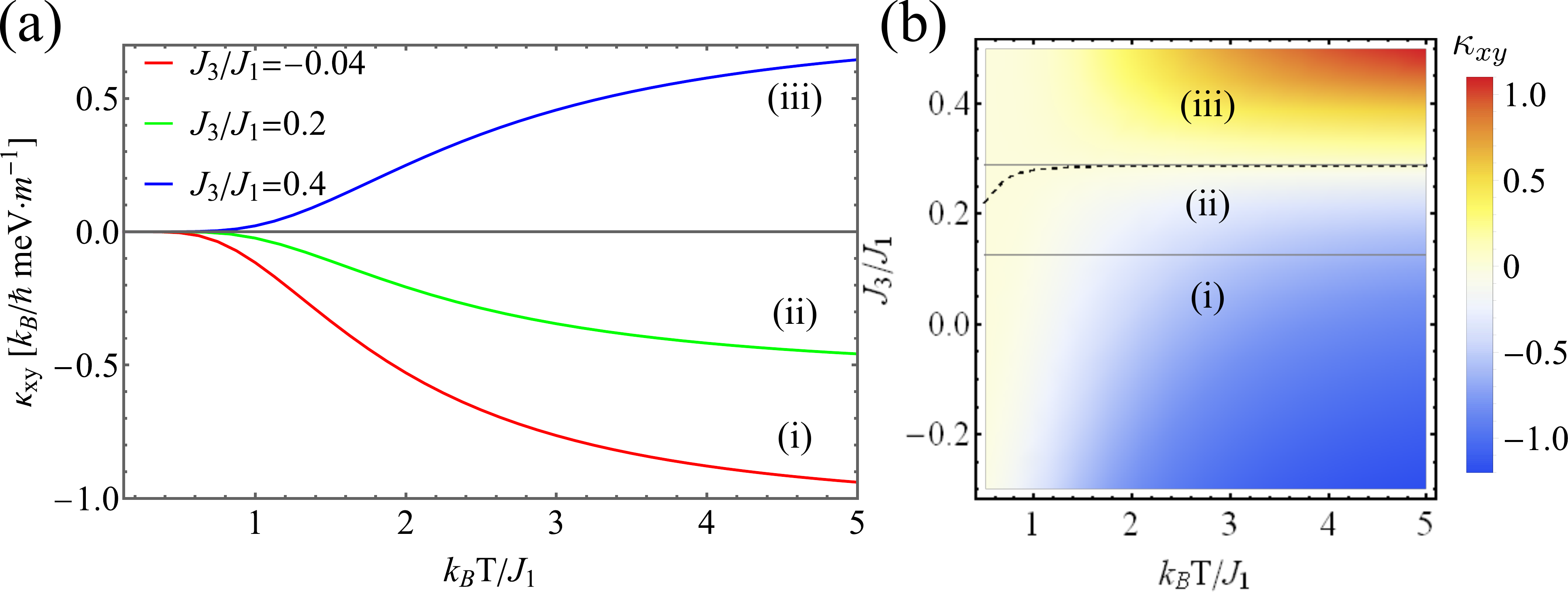}
    \caption{Thermal hall conductivity $\kappa_{xy}$ varying with $k_BT$ (a) at $J_3/J_1=$ $-0.04$, $0.2$, and $0.4$. (b) $\kappa_{xy}$ varies with $k_BT$ and $J_3$ at $J_c=0.59$ meV.}
    \label{vsT}
\end{figure}

The thermal Hall conductivity $\kappa_{xy}$ varies with temperature for different values of $J_3$ is shown in Fig.~\ref{vsT}(a) where $J_c=0.59$ meV is kept constant (experimentally determined for CrI$_3$~\cite{cri3}). The full picture of $\kappa_{xy}$ varying with $T$ and $J_3$ is shown as a color plot in Fig.~\ref{vsT}(b). The black dashed line marks the position of zeros $\kappa_{xy}(T,J_3)=0$ where it changes sign. If we extend this line to $T\rightarrow0$ limit, it arrives at $J_{3L}=J_1/8$. In the limit of high temperatures $T>J_1/k_B\approx$~20K, the sign switch occurs at the value of $J_{3H}/J_1=0.288$. 

The variation of $\kappa_{xy}$ can be divided into three regimes as $T$ grows: \\
(i) $J_3<J_{3L}$, $\kappa_{xy}$ remains negative while its magnitude grows with temperature;\\ (ii) $J_{3L}<J_3<J_{3H}$, $\kappa_{xy}$ first switches its sign to negative, then increases in magnitude; \\(iii) $J_3>J_{3H}$, $\kappa_{xy}$ remains positive and grows.

\begin{figure}[!tbp]
    \hspace*{-0.3cm}
    \includegraphics[width=0.45\textwidth]{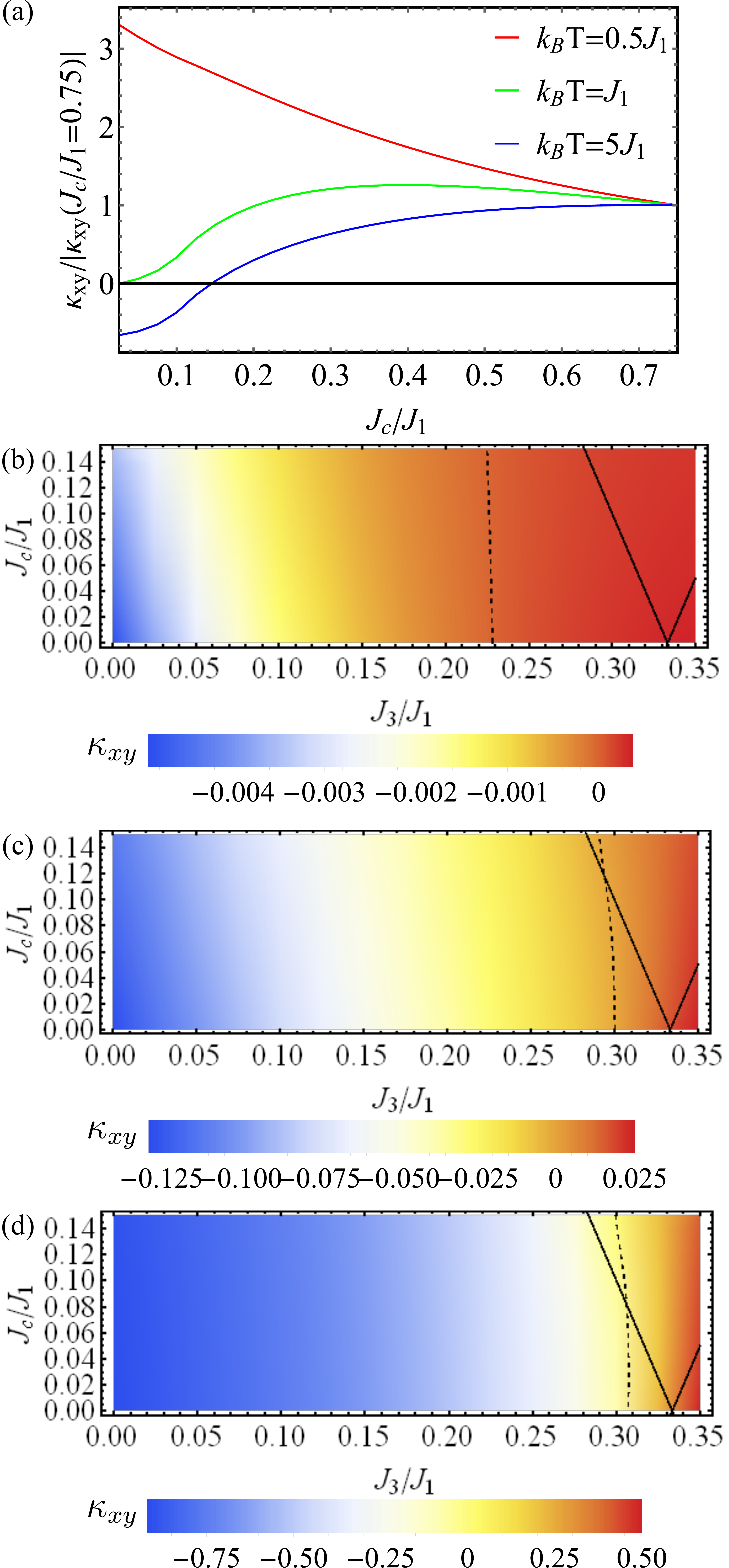}
    \caption{Thermal hall conductivity $\kappa_{xy}$ varying with $J_c$ (a) at $k_BT=0.5J_1$, $J_1$ and $5J_1$. (b)-(d) $\kappa_{xy}$ varies with $J_3$ and $J_c$ at $k_BT=0.5J_1$, $J_1$ and $5J_1$. The dashed lines corresponds to $\kappa_{xy}(J_3,J_c)=0$.}
    \label{vsJc}
\end{figure}

\subsection{Temperature evolution of $\kappa_{xy}$ as a function of interlayer coupling}

Varying the magnitude of the 3rd-neighbor coupling $J_3$ may not be easily achievable in a given compound, however the value of the interlayer coupling $J_c$ should be susceptible to the uniaxial strain applied along the $c$-axis in these van der Waals coupled layered ferromagnets.
We therefore investigate the behaviour of 
$\kappa_{xy}$ varies with $J_c$ shown in Fig.~\ref{vsJc}(a) (while maintaining $J_3/J_1=0.3$). The full picture of $\kappa_{xy}$ varing with both $J_3$ and $J_c$ is shown as color plots in Fig.~\ref{vsJc}(b)-(d) at several temperatures $k_BT=0.5J_1$, $k_BT=J_1$, and $k_BT=5J_1$.  In all cases, the dashed line marks the position of zeros of $\kappa_{xy}$ where it changes sign. 

As temperature approaches 0, this line becomes vertical (i.e. independent of $J_c$) at a fixed value of $J_3=J_1/8$, as was inferred earlier in Sec.~\ref{4} from Eq.~(\ref{eq.kxy-low}). 
However at higher temperatures, a degree of tunability of the sign of $\kappa_{xy}$ can be achieved by varying $J_c$, provided $J_3$ is initially close to the position of the dashed line. Importantly, this line is not fixed but itself moves to the right upon increasing temperature, and this allows a realistic chance of zeroing-in on the sign change of $\kappa_{xy}$ by varying the uniaxial strain and temperature of the material.


\section{Conclusions}

In this work we investigate the topological properties of the spin-wave excitations in the layered honeycomb lattice ferromagnets, motivated in particular by the recent neutron scattering experiment on CrI$_3$~\cite{cri3}. While the presence of anomalous thermal Hall effect due to 2nd-neighbour Dzyaloshinskii-Moriya interaction is well established in the 2D monolayer regime, here we address the effect of the interlayer coupling  in the 3D layered ferromagnets, adopting the ABC stacking of layers realized in CrI$_3$.  
We find that this affects qualitatively the topological properties of the model, resulting most notably in the intermediate \textit{gapless phase} upon increasing the 3rd neighbour coupling $J_3$ in the plane. This gapless phase, which only exists for finite interlayer coupling $J_c$ harbours three pairs of Weyl points where the two magnon branches cross linearly. Each pair of Weyl points carries equal and opposite monopole charges, acting as sources and sinks of the Berry curvature in the reciprocal space. 
In complete analogy with Weyl semimetals, we find that this \textit{Weyl magnon} ``conductor" is an intermediate phase between two topological ``insulating" phases. Unlike the electronic case however, where these phases are the trivial band insulator and the topological Chern insulator, the two magnon branches have non-trivial Chern numbers on either side of the gapless phase: $\mathcal{C}_{-}=+1$ on the left (for $J_3/J_1< 1/3 -\delta_c$) and $\mathcal{C}_{-}=-2$ on the right (for $J_3/J_1> 1/3 +\delta_c$). This sign change of the Berry curvature manifests itself also in the sign of the thermal Hall effect $\kappa_{xy}(T)$, which we compute in different temperature regimes.

In the low-temperature regime $T\ll |J|$, we obtain an analytical result $\kappa_{xy(T\rightarrow0)}\sim T^{\frac{5}{2}}\exp(-\frac{3K}{k_BT})$, where $K$ is the Ising anisotropy constant. In the opposite limit of high temperatures, higher than the magnon bandwidth, $\kappa_{xy}$ approaches a constant value, whose sign switches upon increasing $J_3$ according to the change in sign of the Chern number of the magnon branches, as noted above. Interestingly, the presence of the Weyl points in the intermediate gapless phases goes essentially unnoticed in the low-temperature limit which is dominated by the Berry curvature at the bottom of the dispersion, away from Weyl points. By contrast, at finite temperatures comparable to $|J|$, $\kappa_{xy}$ is sensitive to the development of the Weyl points.

Finally, we investigate the dependence of $\kappa_{xy}$ on the interlayer coupling $J_c$. Similar to a previous work~\cite{owerre2}, $\kappa_{xy}$ will be suppressed while $J_c$ increasing. But a significant difference is that thermal hall effect may change sign as $J_c$ changes. As shown in Fig.\ref{vsJc} (b), the curve marking the position of zeros $\kappa_{xy}(J_3,J_c)=0$ is not vertical at finite temperature, meaning that a constant $J_3$ vertical line  may intersect with it in one point. It means that in the range of $J_3$ where this curve appears, changing the interlayer coupling $J_c$ will lead to a sign switch of $\kappa_{xy}$. This is a new phenomenon which can be experimentally probed by applying a uniaxial strain to the sample along the $c$-axis perpendicular to the layers. Given the van der Waals nature of the coupling between the layers, even a moderate amount of uniaxial strain may result in significant changes of the interlayer spin coupling $J_c$.


\section{Acknowledgements}
This work was supported by the Robert A. Welch grant no. C-1818. The authors acknowledge the hospitality of the Kavli Institute for Theoretical Physics, supported by the National Science Foundation under Grant No. NSF PHY-1748958, where part of this work was performed. 

\appendix
\section{The boundary of gapless phase}\label{app.H}

\begin{figure}[htbp]
    \centering
    \includegraphics[width=0.45\textwidth]{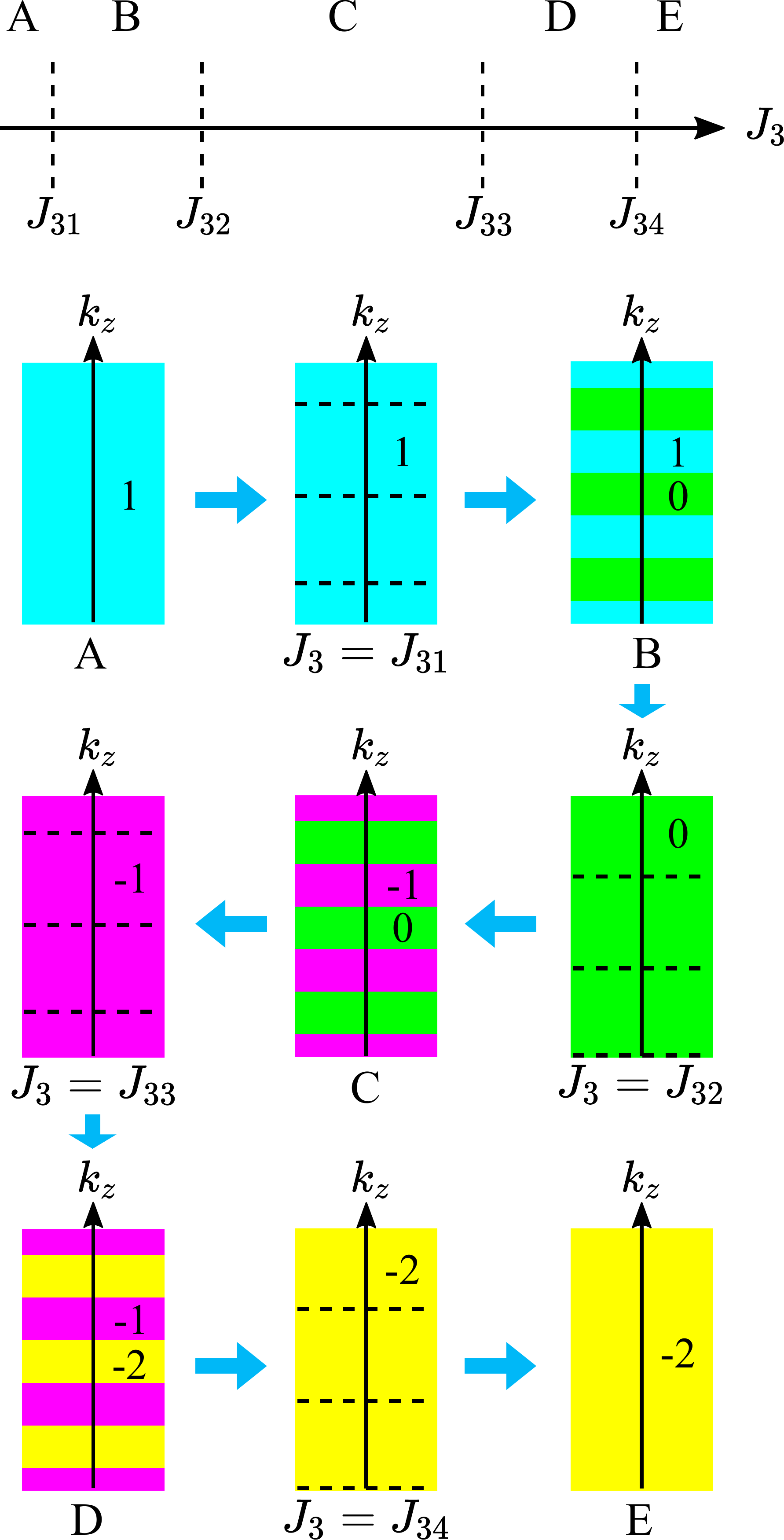}
    \caption{Chern number of lower band $\mathcal{C}_{-}(k_z)$ varying with $k_z$ in different $J_3$ region. In phases B, C and D, the boundaries of different color are $k_z$ planes where gapless points are located. At the critical values $J_3=J_{31},J_{32},J_{33},J_{34}$, the dashed lines correspond to the planes where a pair of gapless points are located at the same value of $k_z$, and their monopole charges cancel out -- as a result, they do not lead to the change of the Chern number.}
    \label{CNchange}
\end{figure}

After the Holstein Primakoff transformation and transforming into momentum space, the Hamiltonian matrix $\mathcal{H}(\vec{k})$ is
\begin{equation}\label{eq.matrixH}
\mathcal{H}(\vec{k})=h_0(\vec{k})\sigma_0+h_x(\vec{k})\sigma_x+h_y(\vec{k})\sigma_y+h_z(\vec{k})\sigma_z,
\end{equation}
where
\begin{equation}
\begin{aligned}
h_0(\vec{k})&=3J_1S+6J_2S+3J_3S+2KS+H_z+J_cS\\
&-2SJ_2(2\cos(\frac{\sqrt3k_x}{2})\cos(\frac{3k_y}{2})+\cos(\sqrt3k_x)),\\
h_x(\vec{k})&=-J_1S(2\cos(\frac{\sqrt3k_x}{2})\cos(\frac{k_y}{2})+\cos(k_y))\\
&-J_3S(2\cos(\sqrt3k_x)\cos(k_y)+\cos(2k_y))-J_cS\cos(ck_z),\\
h_y(\vec{k})&=-J_1S(2\cos(\frac{\sqrt3k_x}{2})\sin(\frac{k_y}{2})-\sin(k_y))\\
&-J_3S(-2\cos(\sqrt3k_x)\sin(k_y)+\sin(2k_y))-J_cS\sin(ck_z),\\
h_z(\vec{k})&=2SD(2\sin(\frac{\sqrt3k_x}{2})(\cos(\frac{\sqrt3k_x}{2})-\cos(\frac{3k_y}{2}))).
\end{aligned}
\end{equation}

In the gapless case $h_x(\vec{k})=h_y(\vec{k})=h_z(\vec{k})=0$. After eliminate $k_z$ by using $h_x(\vec{k})=h_y(\vec{k})=0$, we obtain

\begin{equation}\label{EQ}
\begin{aligned}
    J_ce^{ik_z}=&-J_1(e^{i\frac{1}{2}(\sqrt{3}k_x+k_y)}+e^{i\frac{1}{2}(-\sqrt{3}k_x+k_y)}+e^{-ik_y})\\
    &-J_3(e^{i(\sqrt{3}k_x-k_y)}+e^{i(-\sqrt{3}k_x-k_y)}+e^{i2k_y}).
\end{aligned}
\end{equation}
 Another equation $h_z(\vec{k})=0$ gives
\begin{equation}
    \sin(\frac{\sqrt{3}}{2}k_x)(\cos(\frac{\sqrt{3}}{2}k_x)-\cos\frac{3}{2}k_y)=0.
\end{equation}
In the first BZ, the three solutions are
\begin{equation}
    k_x=0, k_y=\pm\frac{\sqrt{3}}{3}k_x,
\end{equation}
which are obviously equivalent under $120^{\circ}$ rotation symmetry. So we can choose $k_x=0$, take the square of modulus of $J_ce^{ik_z}$, then the equation Eq.(\ref{EQ}) becomes
\begin{equation}\label{eq.solution}
\begin{aligned}
    &(4J_1J_3+8J_3^2)Y^2+(4J_1^2+12J_1J_3)Y\\
    +&(5J_1^2+J_3^2+2J_1J_3-J_c^2)=0,
\end{aligned}
\end{equation}
where $Y=\cos{\frac{3}{2}k_y}$. The solutions are
\begin{equation}
\begin{aligned}
    Y_{1,2}&=\frac{1}{2J_3(J_1+2J_3)}(-J_1^2-3J_1J_3\pm(J_1^4+J_1^3J_3-3J_1^2J_3^2\\
    &-5J_1J_3^3-2J_3^4+J_1J_3J_c^2+2J_3^2J_c^2)^{\frac{1}{2}}).
\end{aligned}    
\end{equation}
The constrains of coupling constants are $-1<J_3/J_1<1$ and $0<J_c/J_1<1$. And we need at least one solution in the range $(-1,1)$. Combination of above equations and inequations gives 
\begin{equation}
\begin{aligned}
&\text{(1)}~0<\frac{J_c}{J_1}\le\frac{1}{8}(-5+3\sqrt{17}),~-1<\frac{J_3}{J_1}<-1+\frac{J_c}{3J_1};\\
&\text{(2)}~0<\frac{J_c}{J_1}\le\frac{1}{8}(-5+3\sqrt{17}),~\frac{1}{3}-\frac{J_c}{3J_1}<\frac{J_3}{J_1}<\frac{1}{3}+\frac{J_c}{3J_1};\\
&\text{(3)}~\frac{1}{8}(-5+3\sqrt{17})<\frac{J_c}{J_1}<1,~\frac{1}{3}-\frac{J_c}{3J_1}<\frac{J_3}{J_1}<x^{*},
\end{aligned}
\end{equation}
where $(-5+3\sqrt{17})/8\approx0.921$ and $x^{*}$ is the second root of
\begin{equation}
    2x^4+5x^3+(3-2J_c^2)x^2+(-1-J_c^2)x-1=0.
\end{equation}
Because usually $J_3$ and $J_c$ are much smaller than $J_1$, we can only consider the solution
\begin{equation}\label{eq.gaplessphase}
    \frac{1}{3}-\frac{J_c}{3J_1}<\frac{J_3}{J_1}<\frac{1}{3}+\frac{J_c}{3J_1}.
\end{equation}
The boundary of gapless phase are given by $J_{31}/J_1=1/3-J_c/3J_1$ and $J_{34}/J_1=1/3+J_c/3J_1$. What's more, if we set $J_c=0$, this solution becomes $J_1=3J_3$, which is the same as the gapless case of monolayer model.

\section{The behaviour of Weyl points with $J_3$ changing} \label{app.Weyl}
After solving the equations of gapless condition $h_x(\vec{k})=h_y(\vec{k})=h_z(\vec{k})=0$, there are 6 gapless points with $k_z=$ $\pm k_{z0}$, $\pm (-k_{z0}+2\pi/3)$ and $\pm (k_{z0}+2\pi/3)$ for $0\le k_{z0}\le \pi/3$. In gapless phase, $\mathcal{C}_{-}(k_z)$ changes to $\mathcal{C}_{-}(k_z)\pm1$ across the $k_z$ planes which contains one gapless point. Analogous to Weyl semimetal, these pairs of gapless points are Weyl points. If two Weyl points with opposite charge live in the same $k_z$ plane, $\mathcal{C}_{-}(k_z)$ keeps unchanged after cross it. This case appears at $k_{z0}=0$ and $\pi/3$. Bring it back into $h_x(\vec{k})=h_y(\vec{k})=h_z(\vec{k})=0$ it gives the phase boundaries $J_3=J_{32},J_{33}$ in gapless phase.

As $J_3$ increasing, we can devide the process into 5 topological phases:
\begin{itemize}
  \item $J_3<J_{31}$, the system is gapped, $\mathcal{C}_-(k_z)=+1$;
  \item $J_{31}<J_3<J_{32}$, the system is gapless, 6 planes with Weyl points separates 6 regions alternating with $\mathcal{C}_-(k_z)=+1$ and $\mathcal{C}_-(k_z)=0$;
  \item $J_{32}<J_3<J_{33}$, the system is gapless, 6 planes with Weyl points separates 6 regions alternating with $\mathcal{C}_-(k_z)=-1$ and $\mathcal{C}_-(k_z)=0$;
  \item $J_{33}<J_3<J_{34}$, the system is gapless, 6 planes with Weyl points separates 6 regions alternating with $\mathcal{C}_-(k_z)=-1$ and $\mathcal{C}_-(k_z)=-2$;
  \item $J_{34}<J_3$, the system is gapped, $\mathcal{C}_-(k_z)=-2$.
\end{itemize}

Fig.~\ref{CNchange} shows the variation of lower band Chern number $\mathcal{C}_-(k_z)$ along $k_z$ with increasing $J_3$.

\bibliography{cite}

\begin{thebibliography}{58}
\expandafter\ifx\csname natexlab\endcsname\relax\def\natexlab#1{#1}\fi
\expandafter\ifx\csname bibnamefont\endcsname\relax
  \def\bibnamefont#1{#1}\fi
\expandafter\ifx\csname bibfnamefont\endcsname\relax
  \def\bibfnamefont#1{#1}\fi
\expandafter\ifx\csname citenamefont\endcsname\relax
  \def\citenamefont#1{#1}\fi
\expandafter\ifx\csname url\endcsname\relax
  \def\url#1{\texttt{#1}}\fi
\expandafter\ifx\csname urlprefix\endcsname\relax\def\urlprefix{URL }\fi
\providecommand{\bibinfo}[2]{#2}
\providecommand{\eprint}[2][]{\url{#2}}

\bibitem[{\citenamefont{Bloch}(1930)}]{magnon1}
\bibinfo{author}{\bibfnamefont{F.}~\bibnamefont{Bloch}},
  \bibinfo{journal}{Zeitschrift f{\"u}r Physik} \textbf{\bibinfo{volume}{61}},
  \bibinfo{pages}{206} (\bibinfo{year}{1930}).

\bibitem[{\citenamefont{Holstein and Primakoff}(1940)}]{magnon2}
\bibinfo{author}{\bibfnamefont{T.}~\bibnamefont{Holstein}} \bibnamefont{and}
  \bibinfo{author}{\bibfnamefont{H.}~\bibnamefont{Primakoff}},
  \bibinfo{journal}{Phys. Rev.} \textbf{\bibinfo{volume}{58}},
  \bibinfo{pages}{1098} (\bibinfo{year}{1940}),
  \urlprefix\url{https://link.aps.org/doi/10.1103/PhysRev.58.1098}.

\bibitem[{\citenamefont{Majlis}(2007)}]{magnon3}
\bibinfo{author}{\bibfnamefont{N.}~\bibnamefont{Majlis}},
  \emph{\bibinfo{title}{The quantum theory of magnetism}}
  (\bibinfo{publisher}{World Scientific Publishing Company},
  \bibinfo{year}{2007}).

\bibitem[{\citenamefont{Demokritov et~al.}(2006)\citenamefont{Demokritov,
  Demidov, Dzyapko, Melkov, Serga, Hillebrands, and Slavin}}]{magnon4}
\bibinfo{author}{\bibfnamefont{S.~O.} \bibnamefont{Demokritov}},
  \bibinfo{author}{\bibfnamefont{V.~E.} \bibnamefont{Demidov}},
  \bibinfo{author}{\bibfnamefont{O.}~\bibnamefont{Dzyapko}},
  \bibinfo{author}{\bibfnamefont{G.~A.} \bibnamefont{Melkov}},
  \bibinfo{author}{\bibfnamefont{A.~A.} \bibnamefont{Serga}},
  \bibinfo{author}{\bibfnamefont{B.}~\bibnamefont{Hillebrands}},
  \bibnamefont{and} \bibinfo{author}{\bibfnamefont{A.~N.}
  \bibnamefont{Slavin}}, \bibinfo{journal}{Nature}
  \textbf{\bibinfo{volume}{443}}, \bibinfo{pages}{430} (\bibinfo{year}{2006}).

\bibitem[{\citenamefont{Serga et~al.}(2014)\citenamefont{Serga, Tiberkevich,
  Sandweg, Vasyuchka, Bozhko, Chumak, Neumann, Obry, Melkov, Slavin
  et~al.}}]{magnon5}
\bibinfo{author}{\bibfnamefont{A.~A.} \bibnamefont{Serga}},
  \bibinfo{author}{\bibfnamefont{V.~S.} \bibnamefont{Tiberkevich}},
  \bibinfo{author}{\bibfnamefont{C.~W.} \bibnamefont{Sandweg}},
  \bibinfo{author}{\bibfnamefont{V.~I.} \bibnamefont{Vasyuchka}},
  \bibinfo{author}{\bibfnamefont{D.~A.} \bibnamefont{Bozhko}},
  \bibinfo{author}{\bibfnamefont{A.~V.} \bibnamefont{Chumak}},
  \bibinfo{author}{\bibfnamefont{T.}~\bibnamefont{Neumann}},
  \bibinfo{author}{\bibfnamefont{B.}~\bibnamefont{Obry}},
  \bibinfo{author}{\bibfnamefont{G.~A.} \bibnamefont{Melkov}},
  \bibinfo{author}{\bibfnamefont{A.~N.} \bibnamefont{Slavin}},
  \bibnamefont{et~al.}, \bibinfo{journal}{Nature communications}
  \textbf{\bibinfo{volume}{5}}, \bibinfo{pages}{1} (\bibinfo{year}{2014}).

\bibitem[{\citenamefont{Clausen et~al.}(2015)\citenamefont{Clausen, Bozhko,
  Vasyuchka, Hillebrands, Melkov, and Serga}}]{magnon6}
\bibinfo{author}{\bibfnamefont{P.}~\bibnamefont{Clausen}},
  \bibinfo{author}{\bibfnamefont{D.~A.} \bibnamefont{Bozhko}},
  \bibinfo{author}{\bibfnamefont{V.~I.} \bibnamefont{Vasyuchka}},
  \bibinfo{author}{\bibfnamefont{B.}~\bibnamefont{Hillebrands}},
  \bibinfo{author}{\bibfnamefont{G.~A.} \bibnamefont{Melkov}},
  \bibnamefont{and} \bibinfo{author}{\bibfnamefont{A.~A.} \bibnamefont{Serga}},
  \bibinfo{journal}{Phys. Rev. B} \textbf{\bibinfo{volume}{91}},
  \bibinfo{pages}{220402} (\bibinfo{year}{2015}),
  \urlprefix\url{https://link.aps.org/doi/10.1103/PhysRevB.91.220402}.

\bibitem[{\citenamefont{Bozhko et~al.}(2016)\citenamefont{Bozhko, Serga,
  Clausen, Vasyuchka, Heussner, Melkov, Pomyalov, L’vov, and
  Hillebrands}}]{magnon7}
\bibinfo{author}{\bibfnamefont{D.~A.} \bibnamefont{Bozhko}},
  \bibinfo{author}{\bibfnamefont{A.~A.} \bibnamefont{Serga}},
  \bibinfo{author}{\bibfnamefont{P.}~\bibnamefont{Clausen}},
  \bibinfo{author}{\bibfnamefont{V.~I.} \bibnamefont{Vasyuchka}},
  \bibinfo{author}{\bibfnamefont{F.}~\bibnamefont{Heussner}},
  \bibinfo{author}{\bibfnamefont{G.~A.} \bibnamefont{Melkov}},
  \bibinfo{author}{\bibfnamefont{A.}~\bibnamefont{Pomyalov}},
  \bibinfo{author}{\bibfnamefont{V.~S.} \bibnamefont{L’vov}},
  \bibnamefont{and}
  \bibinfo{author}{\bibfnamefont{B.}~\bibnamefont{Hillebrands}},
  \bibinfo{journal}{Nature Physics} \textbf{\bibinfo{volume}{12}},
  \bibinfo{pages}{1057} (\bibinfo{year}{2016}).

\bibitem[{\citenamefont{Bennemann and Ketterson}(2013)}]{magnon8}
\bibinfo{author}{\bibfnamefont{K.-H.} \bibnamefont{Bennemann}}
  \bibnamefont{and} \bibinfo{author}{\bibfnamefont{J.~B.}
  \bibnamefont{Ketterson}}, \emph{\bibinfo{title}{Novel superfluids}},
  vol.~\bibinfo{volume}{1} (\bibinfo{publisher}{OUP Oxford},
  \bibinfo{year}{2013}).

\bibitem[{\citenamefont{Kajiwara et~al.}(2010)\citenamefont{Kajiwara, Harii,
  Takahashi, Ohe, Uchida, Mizuguchi, Umezawa, Kawai, Ando, Takanashi
  et~al.}}]{magnon9}
\bibinfo{author}{\bibfnamefont{Y.}~\bibnamefont{Kajiwara}},
  \bibinfo{author}{\bibfnamefont{K.}~\bibnamefont{Harii}},
  \bibinfo{author}{\bibfnamefont{S.}~\bibnamefont{Takahashi}},
  \bibinfo{author}{\bibfnamefont{J.-i.} \bibnamefont{Ohe}},
  \bibinfo{author}{\bibfnamefont{K.}~\bibnamefont{Uchida}},
  \bibinfo{author}{\bibfnamefont{M.}~\bibnamefont{Mizuguchi}},
  \bibinfo{author}{\bibfnamefont{H.}~\bibnamefont{Umezawa}},
  \bibinfo{author}{\bibfnamefont{H.}~\bibnamefont{Kawai}},
  \bibinfo{author}{\bibfnamefont{K.}~\bibnamefont{Ando}},
  \bibinfo{author}{\bibfnamefont{K.}~\bibnamefont{Takanashi}},
  \bibnamefont{et~al.}, \bibinfo{journal}{Nature}
  \textbf{\bibinfo{volume}{464}}, \bibinfo{pages}{262} (\bibinfo{year}{2010}).

\bibitem[{\citenamefont{Fransson et~al.}(2016)\citenamefont{Fransson,
  Black-Schaffer, and Balatsky}}]{diracmg1}
\bibinfo{author}{\bibfnamefont{J.}~\bibnamefont{Fransson}},
  \bibinfo{author}{\bibfnamefont{A.~M.} \bibnamefont{Black-Schaffer}},
  \bibnamefont{and} \bibinfo{author}{\bibfnamefont{A.~V.}
  \bibnamefont{Balatsky}}, \bibinfo{journal}{Phys. Rev. B}
  \textbf{\bibinfo{volume}{94}}, \bibinfo{pages}{075401}
  (\bibinfo{year}{2016}), ISSN \bibinfo{issn}{2469-9950, 2469-9969},
  \urlprefix\url{https://link.aps.org/doi/10.1103/PhysRevB.94.075401}.

\bibitem[{\citenamefont{Pershoguba et~al.}(2018)\citenamefont{Pershoguba,
  Banerjee, Lashley, Park, {\AA}gren, Aeppli, and Balatsky}}]{diracmg2}
\bibinfo{author}{\bibfnamefont{S.~S.} \bibnamefont{Pershoguba}},
  \bibinfo{author}{\bibfnamefont{S.}~\bibnamefont{Banerjee}},
  \bibinfo{author}{\bibfnamefont{J.}~\bibnamefont{Lashley}},
  \bibinfo{author}{\bibfnamefont{J.}~\bibnamefont{Park}},
  \bibinfo{author}{\bibfnamefont{H.}~\bibnamefont{{\AA}gren}},
  \bibinfo{author}{\bibfnamefont{G.}~\bibnamefont{Aeppli}}, \bibnamefont{and}
  \bibinfo{author}{\bibfnamefont{A.~V.} \bibnamefont{Balatsky}},
  \bibinfo{journal}{Physical Review X} \textbf{\bibinfo{volume}{8}},
  \bibinfo{pages}{011010} (\bibinfo{year}{2018}).

\bibitem[{\citenamefont{Haldane}(1988)}]{Haldane}
\bibinfo{author}{\bibfnamefont{F.~D.~M.} \bibnamefont{Haldane}},
  \bibinfo{journal}{Phys. Rev. Lett.} \textbf{\bibinfo{volume}{61}},
  \bibinfo{pages}{2015} (\bibinfo{year}{1988}), \bibinfo{note}{publisher:
  American Physical Society},
  \urlprefix\url{https://link.aps.org/doi/10.1103/PhysRevLett.61.2015}.

\bibitem[{\citenamefont{Owerre}(2016{\natexlab{a}})}]{owerre1}
\bibinfo{author}{\bibfnamefont{S.~A.} \bibnamefont{Owerre}},
  \bibinfo{journal}{Journal of Physics: Condensed Matter}
  \textbf{\bibinfo{volume}{28}}, \bibinfo{pages}{386001}
  (\bibinfo{year}{2016}{\natexlab{a}}),
  \urlprefix\url{https://doi.org/10.1088%2F0953-8984%2F28%2F38%2F386001}.

\bibitem[{\citenamefont{Owerre}(2016{\natexlab{b}})}]{owerre2}
\bibinfo{author}{\bibfnamefont{S.~A.} \bibnamefont{Owerre}},
  \bibinfo{journal}{Journal of Applied Physics} \textbf{\bibinfo{volume}{120}},
  \bibinfo{pages}{043903} (\bibinfo{year}{2016}{\natexlab{b}}),
  \eprint{https://doi.org/10.1063/1.4959815},
  \urlprefix\url{https://doi.org/10.1063/1.4959815}.

\bibitem[{\citenamefont{Lee et~al.}(2018)\citenamefont{Lee, Chung, Park, and
  Park}}]{honey1}
\bibinfo{author}{\bibfnamefont{K.~H.} \bibnamefont{Lee}},
  \bibinfo{author}{\bibfnamefont{S.~B.} \bibnamefont{Chung}},
  \bibinfo{author}{\bibfnamefont{K.}~\bibnamefont{Park}}, \bibnamefont{and}
  \bibinfo{author}{\bibfnamefont{J.-G.} \bibnamefont{Park}},
  \bibinfo{journal}{Phys. Rev. B} \textbf{\bibinfo{volume}{97}},
  \bibinfo{pages}{180401} (\bibinfo{year}{2018}),
  \urlprefix\url{https://link.aps.org/doi/10.1103/PhysRevB.97.180401}.

\bibitem[{\citenamefont{Dzyaloshinsky}(1958)}]{DM1}
\bibinfo{author}{\bibfnamefont{I.}~\bibnamefont{Dzyaloshinsky}},
  \bibinfo{journal}{Journal of Physics and Chemistry of Solids}
  \textbf{\bibinfo{volume}{4}}, \bibinfo{pages}{241 } (\bibinfo{year}{1958}),
  ISSN \bibinfo{issn}{0022-3697},
  \urlprefix\url{http://www.sciencedirect.com/science/article/pii/0022369758900763}.

\bibitem[{\citenamefont{Moriya}(1960)}]{DM2}
\bibinfo{author}{\bibfnamefont{T.}~\bibnamefont{Moriya}},
  \bibinfo{journal}{Phys. Rev.} \textbf{\bibinfo{volume}{120}},
  \bibinfo{pages}{91} (\bibinfo{year}{1960}),
  \urlprefix\url{https://link.aps.org/doi/10.1103/PhysRev.120.91}.

\bibitem[{\citenamefont{Katsura et~al.}(2010)\citenamefont{Katsura, Nagaosa,
  and Lee}}]{MagnonHEth1}
\bibinfo{author}{\bibfnamefont{H.}~\bibnamefont{Katsura}},
  \bibinfo{author}{\bibfnamefont{N.}~\bibnamefont{Nagaosa}}, \bibnamefont{and}
  \bibinfo{author}{\bibfnamefont{P.~A.} \bibnamefont{Lee}},
  \bibinfo{journal}{Phys. Rev. Lett.} \textbf{\bibinfo{volume}{104}},
  \bibinfo{pages}{066403} (\bibinfo{year}{2010}),
  \urlprefix\url{https://link.aps.org/doi/10.1103/PhysRevLett.104.066403}.

\bibitem[{\citenamefont{Onose et~al.}(2010)\citenamefont{Onose, Ideue, Katsura,
  Shiomi, Nagaosa, and Tokura}}]{Onose297}
\bibinfo{author}{\bibfnamefont{Y.}~\bibnamefont{Onose}},
  \bibinfo{author}{\bibfnamefont{T.}~\bibnamefont{Ideue}},
  \bibinfo{author}{\bibfnamefont{H.}~\bibnamefont{Katsura}},
  \bibinfo{author}{\bibfnamefont{Y.}~\bibnamefont{Shiomi}},
  \bibinfo{author}{\bibfnamefont{N.}~\bibnamefont{Nagaosa}}, \bibnamefont{and}
  \bibinfo{author}{\bibfnamefont{Y.}~\bibnamefont{Tokura}},
  \bibinfo{journal}{Science} \textbf{\bibinfo{volume}{329}},
  \bibinfo{pages}{297} (\bibinfo{year}{2010}), ISSN \bibinfo{issn}{0036-8075},
  \eprint{https://science.sciencemag.org/content/329/5989/297.full.pdf},
  \urlprefix\url{https://science.sciencemag.org/content/329/5989/297}.

\bibitem[{\citenamefont{Matsumoto and
  Murakami}(2011{\natexlab{a}})}]{MagnonHEth2}
\bibinfo{author}{\bibfnamefont{R.}~\bibnamefont{Matsumoto}} \bibnamefont{and}
  \bibinfo{author}{\bibfnamefont{S.}~\bibnamefont{Murakami}},
  \bibinfo{journal}{Phys. Rev. Lett.} \textbf{\bibinfo{volume}{106}},
  \bibinfo{pages}{197202} (\bibinfo{year}{2011}{\natexlab{a}}),
  \urlprefix\url{https://link.aps.org/doi/10.1103/PhysRevLett.106.197202}.

\bibitem[{\citenamefont{Matsumoto and
  Murakami}(2011{\natexlab{b}})}]{MagnonHEth3}
\bibinfo{author}{\bibfnamefont{R.}~\bibnamefont{Matsumoto}} \bibnamefont{and}
  \bibinfo{author}{\bibfnamefont{S.}~\bibnamefont{Murakami}},
  \bibinfo{journal}{Phys. Rev. B} \textbf{\bibinfo{volume}{84}},
  \bibinfo{pages}{184406} (\bibinfo{year}{2011}{\natexlab{b}}),
  \urlprefix\url{https://link.aps.org/doi/10.1103/PhysRevB.84.184406}.

\bibitem[{\citenamefont{Shindou
  et~al.}(2013{\natexlab{a}})\citenamefont{Shindou, Matsumoto, Murakami, and
  Ohe}}]{MagnonHEth4}
\bibinfo{author}{\bibfnamefont{R.}~\bibnamefont{Shindou}},
  \bibinfo{author}{\bibfnamefont{R.}~\bibnamefont{Matsumoto}},
  \bibinfo{author}{\bibfnamefont{S.}~\bibnamefont{Murakami}}, \bibnamefont{and}
  \bibinfo{author}{\bibfnamefont{J.-i.} \bibnamefont{Ohe}},
  \bibinfo{journal}{Phys. Rev. B} \textbf{\bibinfo{volume}{87}},
  \bibinfo{pages}{174427} (\bibinfo{year}{2013}{\natexlab{a}}),
  \urlprefix\url{https://link.aps.org/doi/10.1103/PhysRevB.87.174427}.

\bibitem[{\citenamefont{Shindou
  et~al.}(2013{\natexlab{b}})\citenamefont{Shindou, Ohe, Matsumoto, Murakami,
  and Saitoh}}]{MagnonHEth5}
\bibinfo{author}{\bibfnamefont{R.}~\bibnamefont{Shindou}},
  \bibinfo{author}{\bibfnamefont{J.-i.} \bibnamefont{Ohe}},
  \bibinfo{author}{\bibfnamefont{R.}~\bibnamefont{Matsumoto}},
  \bibinfo{author}{\bibfnamefont{S.}~\bibnamefont{Murakami}}, \bibnamefont{and}
  \bibinfo{author}{\bibfnamefont{E.}~\bibnamefont{Saitoh}},
  \bibinfo{journal}{Phys. Rev. B} \textbf{\bibinfo{volume}{87}},
  \bibinfo{pages}{174402} (\bibinfo{year}{2013}{\natexlab{b}}),
  \urlprefix\url{https://link.aps.org/doi/10.1103/PhysRevB.87.174402}.

\bibitem[{\citenamefont{Matsumoto et~al.}(2014)\citenamefont{Matsumoto,
  Shindou, and Murakami}}]{MagnonHEth6}
\bibinfo{author}{\bibfnamefont{R.}~\bibnamefont{Matsumoto}},
  \bibinfo{author}{\bibfnamefont{R.}~\bibnamefont{Shindou}}, \bibnamefont{and}
  \bibinfo{author}{\bibfnamefont{S.}~\bibnamefont{Murakami}},
  \bibinfo{journal}{Phys. Rev. B} \textbf{\bibinfo{volume}{89}},
  \bibinfo{pages}{054420} (\bibinfo{year}{2014}),
  \urlprefix\url{https://link.aps.org/doi/10.1103/PhysRevB.89.054420}.

\bibitem[{\citenamefont{Lee et~al.}(2015)\citenamefont{Lee, Han, and
  Lee}}]{kagome1}
\bibinfo{author}{\bibfnamefont{H.}~\bibnamefont{Lee}},
  \bibinfo{author}{\bibfnamefont{J.~H.} \bibnamefont{Han}}, \bibnamefont{and}
  \bibinfo{author}{\bibfnamefont{P.~A.} \bibnamefont{Lee}},
  \bibinfo{journal}{Phys. Rev. B} \textbf{\bibinfo{volume}{91}},
  \bibinfo{pages}{125413} (\bibinfo{year}{2015}),
  \urlprefix\url{https://link.aps.org/doi/10.1103/PhysRevB.91.125413}.

\bibitem[{\citenamefont{Mook et~al.}(2014{\natexlab{a}})\citenamefont{Mook,
  Henk, and Mertig}}]{kagome2}
\bibinfo{author}{\bibfnamefont{A.}~\bibnamefont{Mook}},
  \bibinfo{author}{\bibfnamefont{J.}~\bibnamefont{Henk}}, \bibnamefont{and}
  \bibinfo{author}{\bibfnamefont{I.}~\bibnamefont{Mertig}},
  \bibinfo{journal}{Phys. Rev. B} \textbf{\bibinfo{volume}{90}},
  \bibinfo{pages}{024412} (\bibinfo{year}{2014}{\natexlab{a}}),
  \urlprefix\url{https://link.aps.org/doi/10.1103/PhysRevB.90.024412}.

\bibitem[{\citenamefont{Mook et~al.}(2014{\natexlab{b}})\citenamefont{Mook,
  Henk, and Mertig}}]{kagome3}
\bibinfo{author}{\bibfnamefont{A.}~\bibnamefont{Mook}},
  \bibinfo{author}{\bibfnamefont{J.}~\bibnamefont{Henk}}, \bibnamefont{and}
  \bibinfo{author}{\bibfnamefont{I.}~\bibnamefont{Mertig}},
  \bibinfo{journal}{Phys. Rev. B} \textbf{\bibinfo{volume}{89}},
  \bibinfo{pages}{134409} (\bibinfo{year}{2014}{\natexlab{b}}),
  \urlprefix\url{https://link.aps.org/doi/10.1103/PhysRevB.89.134409}.

\bibitem[{\citenamefont{Hirschberger et~al.}(2015)\citenamefont{Hirschberger,
  Chisnell, Lee, and Ong}}]{kagome4}
\bibinfo{author}{\bibfnamefont{M.}~\bibnamefont{Hirschberger}},
  \bibinfo{author}{\bibfnamefont{R.}~\bibnamefont{Chisnell}},
  \bibinfo{author}{\bibfnamefont{Y.~S.} \bibnamefont{Lee}}, \bibnamefont{and}
  \bibinfo{author}{\bibfnamefont{N.~P.} \bibnamefont{Ong}},
  \bibinfo{journal}{Phys. Rev. Lett.} \textbf{\bibinfo{volume}{115}},
  \bibinfo{pages}{106603} (\bibinfo{year}{2015}),
  \urlprefix\url{https://link.aps.org/doi/10.1103/PhysRevLett.115.106603}.

\bibitem[{\citenamefont{Cao et~al.}(2015)\citenamefont{Cao, Chen, and
  He}}]{lieb1}
\bibinfo{author}{\bibfnamefont{X.}~\bibnamefont{Cao}},
  \bibinfo{author}{\bibfnamefont{K.}~\bibnamefont{Chen}}, \bibnamefont{and}
  \bibinfo{author}{\bibfnamefont{D.}~\bibnamefont{He}},
  \bibinfo{journal}{Journal of Physics: Condensed Matter}
  \textbf{\bibinfo{volume}{27}}, \bibinfo{pages}{166003}
  (\bibinfo{year}{2015}),
  \urlprefix\url{https://doi.org/10.1088%2F0953-8984%2F27%2F16%2F166003}.

\bibitem[{\citenamefont{Murakami and Okamoto}(2016)}]{Murakami2016}
\bibinfo{author}{\bibfnamefont{S.}~\bibnamefont{Murakami}} \bibnamefont{and}
  \bibinfo{author}{\bibfnamefont{A.}~\bibnamefont{Okamoto}},
  \bibinfo{journal}{J. Phys. Soc. Jpn.} \textbf{\bibinfo{volume}{86}},
  \bibinfo{pages}{011010} (\bibinfo{year}{2016}), ISSN
  \bibinfo{issn}{0031-9015}, \bibinfo{note}{publisher: The Physical Society of
  Japan}, \urlprefix\url{https://journals.jps.jp/doi/10.7566/JPSJ.86.011010}.

\bibitem[{\citenamefont{Wan et~al.}(2011)\citenamefont{Wan, Turner, Vishwanath,
  and Savrasov}}]{Weyl-Wan2011}
\bibinfo{author}{\bibfnamefont{X.}~\bibnamefont{Wan}},
  \bibinfo{author}{\bibfnamefont{A.~M.} \bibnamefont{Turner}},
  \bibinfo{author}{\bibfnamefont{A.}~\bibnamefont{Vishwanath}},
  \bibnamefont{and} \bibinfo{author}{\bibfnamefont{S.~Y.}
  \bibnamefont{Savrasov}}, \bibinfo{journal}{Phys. Rev. B}
  \textbf{\bibinfo{volume}{83}}, \bibinfo{pages}{205101}
  (\bibinfo{year}{2011}), \bibinfo{note}{publisher: American Physical Society},
  \urlprefix\url{https://link.aps.org/doi/10.1103/PhysRevB.83.205101}.

\bibitem[{\citenamefont{Xu et~al.}(2011)\citenamefont{Xu, Weng, Wang, Dai, and
  Fang}}]{Weyl-Chern-Xu2011}
\bibinfo{author}{\bibfnamefont{G.}~\bibnamefont{Xu}},
  \bibinfo{author}{\bibfnamefont{H.}~\bibnamefont{Weng}},
  \bibinfo{author}{\bibfnamefont{Z.}~\bibnamefont{Wang}},
  \bibinfo{author}{\bibfnamefont{X.}~\bibnamefont{Dai}}, \bibnamefont{and}
  \bibinfo{author}{\bibfnamefont{Z.}~\bibnamefont{Fang}},
  \bibinfo{journal}{Phys. Rev. Lett.} \textbf{\bibinfo{volume}{107}},
  \bibinfo{pages}{186806} (\bibinfo{year}{2011}), \bibinfo{note}{publisher:
  American Physical Society},
  \urlprefix\url{https://link.aps.org/doi/10.1103/PhysRevLett.107.186806}.

\bibitem[{\citenamefont{Yang et~al.}(2011)\citenamefont{Yang, Lu, and
  Ran}}]{Weyl-Chern-Yang2011}
\bibinfo{author}{\bibfnamefont{K.-Y.} \bibnamefont{Yang}},
  \bibinfo{author}{\bibfnamefont{Y.-M.} \bibnamefont{Lu}}, \bibnamefont{and}
  \bibinfo{author}{\bibfnamefont{Y.}~\bibnamefont{Ran}},
  \bibinfo{journal}{Phys. Rev. B} \textbf{\bibinfo{volume}{84}},
  \bibinfo{pages}{075129} (\bibinfo{year}{2011}), \bibinfo{note}{publisher:
  American Physical Society},
  \urlprefix\url{https://link.aps.org/doi/10.1103/PhysRevB.84.075129}.

\bibitem[{\citenamefont{Burkov and Balents}(2011)}]{EFMWeylSM1}
\bibinfo{author}{\bibfnamefont{A.~A.} \bibnamefont{Burkov}} \bibnamefont{and}
  \bibinfo{author}{\bibfnamefont{L.}~\bibnamefont{Balents}},
  \bibinfo{journal}{Phys. Rev. Lett.} \textbf{\bibinfo{volume}{107}},
  \bibinfo{pages}{127205} (\bibinfo{year}{2011}),
  \urlprefix\url{https://link.aps.org/doi/10.1103/PhysRevLett.107.127205}.

\bibitem[{\citenamefont{Yan and Felser}(2017)}]{Weyl-felser}
\bibinfo{author}{\bibfnamefont{B.}~\bibnamefont{Yan}} \bibnamefont{and}
  \bibinfo{author}{\bibfnamefont{C.}~\bibnamefont{Felser}},
  \bibinfo{journal}{Annual Review of Condensed Matter Physics}
  \textbf{\bibinfo{volume}{8}}, \bibinfo{pages}{337} (\bibinfo{year}{2017}),
  \bibinfo{note}{\_eprint:
  https://doi.org/10.1146/annurev-conmatphys-031016-025458},
  \urlprefix\url{https://doi.org/10.1146/annurev-conmatphys-031016-025458}.

\bibitem[{\citenamefont{Jin et~al.}(2018)\citenamefont{Jin, Wang, Xia, Zheng,
  and Xu}}]{EFMWeylSM2}
\bibinfo{author}{\bibfnamefont{Y.~J.} \bibnamefont{Jin}},
  \bibinfo{author}{\bibfnamefont{R.}~\bibnamefont{Wang}},
  \bibinfo{author}{\bibfnamefont{B.~W.} \bibnamefont{Xia}},
  \bibinfo{author}{\bibfnamefont{B.~B.} \bibnamefont{Zheng}}, \bibnamefont{and}
  \bibinfo{author}{\bibfnamefont{H.}~\bibnamefont{Xu}}, \bibinfo{journal}{Phys.
  Rev. B} \textbf{\bibinfo{volume}{98}}, \bibinfo{pages}{081101}
  (\bibinfo{year}{2018}),
  \urlprefix\url{https://link.aps.org/doi/10.1103/PhysRevB.98.081101}.

\bibitem[{\citenamefont{Li et~al.}(2016)\citenamefont{Li, Li, Kim, Balents, Yu,
  and Chen}}]{weyl1}
\bibinfo{author}{\bibfnamefont{F.-Y.} \bibnamefont{Li}},
  \bibinfo{author}{\bibfnamefont{Y.-D.} \bibnamefont{Li}},
  \bibinfo{author}{\bibfnamefont{Y.~B.} \bibnamefont{Kim}},
  \bibinfo{author}{\bibfnamefont{L.}~\bibnamefont{Balents}},
  \bibinfo{author}{\bibfnamefont{Y.}~\bibnamefont{Yu}}, \bibnamefont{and}
  \bibinfo{author}{\bibfnamefont{G.}~\bibnamefont{Chen}},
  \bibinfo{journal}{Nature communications} \textbf{\bibinfo{volume}{7}},
  \bibinfo{pages}{12691} (\bibinfo{year}{2016}).

\bibitem[{\citenamefont{Jian and Nie}(2018)}]{weyl2}
\bibinfo{author}{\bibfnamefont{S.-K.} \bibnamefont{Jian}} \bibnamefont{and}
  \bibinfo{author}{\bibfnamefont{W.}~\bibnamefont{Nie}},
  \bibinfo{journal}{Phys. Rev. B} \textbf{\bibinfo{volume}{97}},
  \bibinfo{pages}{115162} (\bibinfo{year}{2018}),
  \urlprefix\url{https://link.aps.org/doi/10.1103/PhysRevB.97.115162}.

\bibitem[{\citenamefont{Owerre}(2018{\natexlab{a}})}]{weyl3}
\bibinfo{author}{\bibfnamefont{S.~A.} \bibnamefont{Owerre}},
  \bibinfo{journal}{Phys. Rev. B} \textbf{\bibinfo{volume}{97}},
  \bibinfo{pages}{094412} (\bibinfo{year}{2018}{\natexlab{a}}),
  \urlprefix\url{https://link.aps.org/doi/10.1103/PhysRevB.97.094412}.

\bibitem[{\citenamefont{Li et~al.}(2017)\citenamefont{Li, Li, Hu, Li, and
  Fang}}]{weyl4}
\bibinfo{author}{\bibfnamefont{K.}~\bibnamefont{Li}},
  \bibinfo{author}{\bibfnamefont{C.}~\bibnamefont{Li}},
  \bibinfo{author}{\bibfnamefont{J.}~\bibnamefont{Hu}},
  \bibinfo{author}{\bibfnamefont{Y.}~\bibnamefont{Li}}, \bibnamefont{and}
  \bibinfo{author}{\bibfnamefont{C.}~\bibnamefont{Fang}},
  \bibinfo{journal}{Phys. Rev. Lett.} \textbf{\bibinfo{volume}{119}},
  \bibinfo{pages}{247202} (\bibinfo{year}{2017}),
  \urlprefix\url{https://link.aps.org/doi/10.1103/PhysRevLett.119.247202}.

\bibitem[{\citenamefont{Yao et~al.}(2018)\citenamefont{Yao, Li, Wang, Xue, Dan,
  Iida, Kamazawa, Li, Fang, and Li}}]{weyl5}
\bibinfo{author}{\bibfnamefont{W.}~\bibnamefont{Yao}},
  \bibinfo{author}{\bibfnamefont{C.}~\bibnamefont{Li}},
  \bibinfo{author}{\bibfnamefont{L.}~\bibnamefont{Wang}},
  \bibinfo{author}{\bibfnamefont{S.}~\bibnamefont{Xue}},
  \bibinfo{author}{\bibfnamefont{Y.}~\bibnamefont{Dan}},
  \bibinfo{author}{\bibfnamefont{K.}~\bibnamefont{Iida}},
  \bibinfo{author}{\bibfnamefont{K.}~\bibnamefont{Kamazawa}},
  \bibinfo{author}{\bibfnamefont{K.}~\bibnamefont{Li}},
  \bibinfo{author}{\bibfnamefont{C.}~\bibnamefont{Fang}}, \bibnamefont{and}
  \bibinfo{author}{\bibfnamefont{Y.}~\bibnamefont{Li}},
  \bibinfo{journal}{Nature Physics} \textbf{\bibinfo{volume}{14}},
  \bibinfo{pages}{1011} (\bibinfo{year}{2018}).

\bibitem[{\citenamefont{Bao et~al.}(2018)\citenamefont{Bao, Wang, Wang, Cai,
  Li, Ma, Wang, Ran, Dong, Abernathy et~al.}}]{weyl6}
\bibinfo{author}{\bibfnamefont{S.}~\bibnamefont{Bao}},
  \bibinfo{author}{\bibfnamefont{J.}~\bibnamefont{Wang}},
  \bibinfo{author}{\bibfnamefont{W.}~\bibnamefont{Wang}},
  \bibinfo{author}{\bibfnamefont{Z.}~\bibnamefont{Cai}},
  \bibinfo{author}{\bibfnamefont{S.}~\bibnamefont{Li}},
  \bibinfo{author}{\bibfnamefont{Z.}~\bibnamefont{Ma}},
  \bibinfo{author}{\bibfnamefont{D.}~\bibnamefont{Wang}},
  \bibinfo{author}{\bibfnamefont{K.}~\bibnamefont{Ran}},
  \bibinfo{author}{\bibfnamefont{Z.-Y.} \bibnamefont{Dong}},
  \bibinfo{author}{\bibfnamefont{D.}~\bibnamefont{Abernathy}},
  \bibnamefont{et~al.}, \bibinfo{journal}{Nature communications}
  \textbf{\bibinfo{volume}{9}}, \bibinfo{pages}{2591} (\bibinfo{year}{2018}).

\bibitem[{\citenamefont{Owerre}(2018{\natexlab{b}})}]{weyl7}
\bibinfo{author}{\bibfnamefont{S.}~\bibnamefont{Owerre}}, \bibinfo{journal}{EPL
  (Europhysics Letters)} \textbf{\bibinfo{volume}{120}}, \bibinfo{pages}{57002}
  (\bibinfo{year}{2018}{\natexlab{b}}).

\bibitem[{\citenamefont{Hwang et~al.}(2017)\citenamefont{Hwang, Trivedi, and
  Randeria}}]{weyl8}
\bibinfo{author}{\bibfnamefont{K.}~\bibnamefont{Hwang}},
  \bibinfo{author}{\bibfnamefont{N.}~\bibnamefont{Trivedi}}, \bibnamefont{and}
  \bibinfo{author}{\bibfnamefont{M.}~\bibnamefont{Randeria}},
  \bibinfo{journal}{arXiv preprint arXiv:1712.08170}  (\bibinfo{year}{2017}).

\bibitem[{\citenamefont{Su et~al.}(2017)\citenamefont{Su, Wang, and
  Wang}}]{weyl9}
\bibinfo{author}{\bibfnamefont{Y.}~\bibnamefont{Su}},
  \bibinfo{author}{\bibfnamefont{X.}~\bibnamefont{Wang}}, \bibnamefont{and}
  \bibinfo{author}{\bibfnamefont{X.}~\bibnamefont{Wang}},
  \bibinfo{journal}{Physical Review B} \textbf{\bibinfo{volume}{95}},
  \bibinfo{pages}{224403} (\bibinfo{year}{2017}).

\bibitem[{\citenamefont{Su and Wang}(2017)}]{weyl11}
\bibinfo{author}{\bibfnamefont{Y.}~\bibnamefont{Su}} \bibnamefont{and}
  \bibinfo{author}{\bibfnamefont{X.}~\bibnamefont{Wang}},
  \bibinfo{journal}{Physical Review B} \textbf{\bibinfo{volume}{96}},
  \bibinfo{pages}{104437} (\bibinfo{year}{2017}).

\bibitem[{\citenamefont{Li and Hu}(2017)}]{weyl10}
\bibinfo{author}{\bibfnamefont{K.-K.} \bibnamefont{Li}} \bibnamefont{and}
  \bibinfo{author}{\bibfnamefont{J.-P.} \bibnamefont{Hu}},
  \bibinfo{journal}{Chinese Physics Letters} \textbf{\bibinfo{volume}{34}},
  \bibinfo{pages}{077501} (\bibinfo{year}{2017}).

\bibitem[{\citenamefont{Zyuzin and Kovalev}(2018)}]{weyl12}
\bibinfo{author}{\bibfnamefont{V.~A.} \bibnamefont{Zyuzin}} \bibnamefont{and}
  \bibinfo{author}{\bibfnamefont{A.~A.} \bibnamefont{Kovalev}},
  \bibinfo{journal}{Physical Review B} \textbf{\bibinfo{volume}{97}},
  \bibinfo{pages}{174407} (\bibinfo{year}{2018}).

\bibitem[{\citenamefont{Mook et~al.}(2016)\citenamefont{Mook, Henk, and
  Mertig}}]{weyl13}
\bibinfo{author}{\bibfnamefont{A.}~\bibnamefont{Mook}},
  \bibinfo{author}{\bibfnamefont{J.}~\bibnamefont{Henk}}, \bibnamefont{and}
  \bibinfo{author}{\bibfnamefont{I.}~\bibnamefont{Mertig}},
  \bibinfo{journal}{Phys. Rev. Lett.} \textbf{\bibinfo{volume}{117}},
  \bibinfo{pages}{157204} (\bibinfo{year}{2016}),
  \urlprefix\url{https://link.aps.org/doi/10.1103/PhysRevLett.117.157204}.

\bibitem[{\citenamefont{Chen et~al.}(2018)\citenamefont{Chen, Chung, Gao, Chen,
  Stone, Kolesnikov, Huang, and Dai}}]{cri3}
\bibinfo{author}{\bibfnamefont{L.}~\bibnamefont{Chen}},
  \bibinfo{author}{\bibfnamefont{J.-H.} \bibnamefont{Chung}},
  \bibinfo{author}{\bibfnamefont{B.}~\bibnamefont{Gao}},
  \bibinfo{author}{\bibfnamefont{T.}~\bibnamefont{Chen}},
  \bibinfo{author}{\bibfnamefont{M.~B.} \bibnamefont{Stone}},
  \bibinfo{author}{\bibfnamefont{A.~I.} \bibnamefont{Kolesnikov}},
  \bibinfo{author}{\bibfnamefont{Q.}~\bibnamefont{Huang}}, \bibnamefont{and}
  \bibinfo{author}{\bibfnamefont{P.}~\bibnamefont{Dai}},
  \bibinfo{journal}{Phys. Rev. X} \textbf{\bibinfo{volume}{8}},
  \bibinfo{pages}{041028} (\bibinfo{year}{2018}),
  \urlprefix\url{https://link.aps.org/doi/10.1103/PhysRevX.8.041028}.

\bibitem[{\citenamefont{Wang et~al.}(2011)\citenamefont{Wang, Eyert, and
  Schwingenschl{\"o}gl}}]{cri3structure1}
\bibinfo{author}{\bibfnamefont{H.}~\bibnamefont{Wang}},
  \bibinfo{author}{\bibfnamefont{V.}~\bibnamefont{Eyert}}, \bibnamefont{and}
  \bibinfo{author}{\bibfnamefont{U.}~\bibnamefont{Schwingenschl{\"o}gl}},
  \bibinfo{journal}{Journal of Physics: Condensed Matter}
  \textbf{\bibinfo{volume}{23}}, \bibinfo{pages}{116003}
  (\bibinfo{year}{2011}).

\bibitem[{\citenamefont{McGuire et~al.}(2015)\citenamefont{McGuire, Dixit,
  Cooper, and Sales}}]{cri3structure2}
\bibinfo{author}{\bibfnamefont{M.~A.} \bibnamefont{McGuire}},
  \bibinfo{author}{\bibfnamefont{H.}~\bibnamefont{Dixit}},
  \bibinfo{author}{\bibfnamefont{V.~R.} \bibnamefont{Cooper}},
  \bibnamefont{and} \bibinfo{author}{\bibfnamefont{B.~C.} \bibnamefont{Sales}},
  \bibinfo{journal}{Chemistry of Materials} \textbf{\bibinfo{volume}{27}},
  \bibinfo{pages}{612} (\bibinfo{year}{2015}).

\bibitem[{\citenamefont{Sivadas et~al.}(2018)\citenamefont{Sivadas, Okamoto,
  Xu, Fennie, and Xiao}}]{cri3structure3}
\bibinfo{author}{\bibfnamefont{N.}~\bibnamefont{Sivadas}},
  \bibinfo{author}{\bibfnamefont{S.}~\bibnamefont{Okamoto}},
  \bibinfo{author}{\bibfnamefont{X.}~\bibnamefont{Xu}},
  \bibinfo{author}{\bibfnamefont{C.~J.} \bibnamefont{Fennie}},
  \bibnamefont{and} \bibinfo{author}{\bibfnamefont{D.}~\bibnamefont{Xiao}},
  \bibinfo{journal}{Nano letters} \textbf{\bibinfo{volume}{18}},
  \bibinfo{pages}{7658} (\bibinfo{year}{2018}).

\bibitem[{\citenamefont{Owerre}(2018{\natexlab{c}})}]{owerre3d}
\bibinfo{author}{\bibfnamefont{S.}~\bibnamefont{Owerre}},
  \bibinfo{journal}{arXiv preprint arXiv:1811.01946}
  (\bibinfo{year}{2018}{\natexlab{c}}).

\bibitem[{\citenamefont{Lin et~al.}(2017)\citenamefont{Lin, Zhuang, Luo, Liu,
  Chen, Yan, Sun, Zhou, Lu, Tong et~al.}}]{CGT-Lin2017}
\bibinfo{author}{\bibfnamefont{G.~T.} \bibnamefont{Lin}},
  \bibinfo{author}{\bibfnamefont{H.~L.} \bibnamefont{Zhuang}},
  \bibinfo{author}{\bibfnamefont{X.}~\bibnamefont{Luo}},
  \bibinfo{author}{\bibfnamefont{B.~J.} \bibnamefont{Liu}},
  \bibinfo{author}{\bibfnamefont{F.~C.} \bibnamefont{Chen}},
  \bibinfo{author}{\bibfnamefont{J.}~\bibnamefont{Yan}},
  \bibinfo{author}{\bibfnamefont{Y.}~\bibnamefont{Sun}},
  \bibinfo{author}{\bibfnamefont{J.}~\bibnamefont{Zhou}},
  \bibinfo{author}{\bibfnamefont{W.~J.} \bibnamefont{Lu}},
  \bibinfo{author}{\bibfnamefont{P.}~\bibnamefont{Tong}}, \bibnamefont{et~al.},
  \bibinfo{journal}{Phys. Rev. B} \textbf{\bibinfo{volume}{95}},
  \bibinfo{pages}{245212} (\bibinfo{year}{2017}), \bibinfo{note}{publisher:
  American Physical Society},
  \urlprefix\url{https://link.aps.org/doi/10.1103/PhysRevB.95.245212}.

\bibitem[{\citenamefont{Weyl}(1929)}]{Weyl1929}
\bibinfo{author}{\bibfnamefont{H.}~\bibnamefont{Weyl}}, \bibinfo{journal}{Z.
  Physik} \textbf{\bibinfo{volume}{56}}, \bibinfo{pages}{330}
  (\bibinfo{year}{1929}), ISSN \bibinfo{issn}{0044-3328},
  \urlprefix\url{https://doi.org/10.1007/BF01339504}.

\bibitem[{\citenamefont{Nakata et~al.}(2017)\citenamefont{Nakata, Klinovaja,
  and Loss}}]{THE-Loss2017}
\bibinfo{author}{\bibfnamefont{K.}~\bibnamefont{Nakata}},
  \bibinfo{author}{\bibfnamefont{J.}~\bibnamefont{Klinovaja}},
  \bibnamefont{and} \bibinfo{author}{\bibfnamefont{D.}~\bibnamefont{Loss}},
  \bibinfo{journal}{Phys. Rev. B} \textbf{\bibinfo{volume}{95}},
  \bibinfo{pages}{125429} (\bibinfo{year}{2017}), \bibinfo{note}{publisher:
  American Physical Society},
  \urlprefix\url{https://link.aps.org/doi/10.1103/PhysRevB.95.125429}.

\bibitem[{\citenamefont{Ye et~al.}(2018)\citenamefont{Ye, Halász, Savary, and
  Balents}}]{CSL-Savary2018}
\bibinfo{author}{\bibfnamefont{M.}~\bibnamefont{Ye}},
  \bibinfo{author}{\bibfnamefont{G.~B.} \bibnamefont{Halász}},
  \bibinfo{author}{\bibfnamefont{L.}~\bibnamefont{Savary}}, \bibnamefont{and}
  \bibinfo{author}{\bibfnamefont{L.}~\bibnamefont{Balents}},
  \bibinfo{journal}{Phys. Rev. Lett.} \textbf{\bibinfo{volume}{121}},
  \bibinfo{pages}{147201} (\bibinfo{year}{2018}), \bibinfo{note}{publisher:
  American Physical Society},
  \urlprefix\url{https://link.aps.org/doi/10.1103/PhysRevLett.121.147201}.

\end{thebibliography}

\end{document}